\documentclass[natbib, graybox, nosecnum]{svmult}

\usepackage{mathptmx}       % selects Times Roman as basic font
\usepackage{helvet}         % selects Helvetica as sans-serif font
\usepackage{courier}        % selects Courier as typewriter font
\usepackage{type1cm}        % activate if the above 3 fonts are
\usepackage{makeidx}         % allows index generation
\usepackage{graphicx}        % standard LaTeX graphics tool
\usepackage{multicol}        % used for the two-column index
\usepackage[bottom]{footmisc}% places footnotes at page bottom
\usepackage{hyperref}        %for hyperlinks
\usepackage{soul}            % for high-lighting of text
\usepackage{tabularx}
\usepackage{amsmath}

\addtocounter{secnumdepth}{1} % adds one level of numbering the sections or paragraphs

\makeindex             % used for the subject index
\begin{document}
%\tableofcontents{}
\title*{Nuclei Near and at the Proton Dripline}
% Use \titlerunning{Short Title} for an abbreviated version of
% your contribution title if the original one is too long
\author{Marek Pf\"utzner \thanks{corresponding author} and Chiara Mazzocchi}
% Use \authorrunning{Short Title} for an abbreviated version of
% your contribution title if the original one is too long
\institute{C. Mazzocchi \at Faculty of Physics, University of Warsaw, Pasteura 5, 02-093
Warszawa, Poland \email{chiara.mazzocchi@fuw.edu.pl}
\and
M. Pf\"utzner \at Faculty of Physics, University of Warsaw, Pasteura 5, 02-093 Warszawa,
Poland, \email{pfutzner@fuw.edu.pl}}
%
% Use the package "url.sty" to avoid
% problems with special characters
% used in your e-mail or web address
%
\maketitle
\abstract{Nuclei in the vicinity of the proton dripline and beyond it are a fascinating realm
within the chart of nuclei. In this chapter the main phenomena that characterise
this domain %isotopes
and are not to be found elsewhere are explored. While moving away from the $\beta$-stability valley towards
the proton dripline, phenomena like very exotic decay modes such as
$\beta$-delayed (multi-) particle emission, and proton-, or two-proton radioactivity are encountered. Landmark
nuclei are here two isotopes with magic proton and neutron numbers, $^{48}$Ni and $^{100}$Sn.
Moreover, proton-rich nuclei ($N < Z$) display other interesting features, like breaking of
isospin symmetry with consequent asymmetry in the energy spectra between mirror nuclei, the
so-called Thomas-Ehrmann shift, or the phenomenon of proton-halo.
%which is due to a very low binding energy of one proton that is able to tunnel far away from
%the nuclear binding potential, resulting in a large spatial extension.
Last but not least, nuclei close to the proton dripline play a very important role in
nucleosynthesis, since they take part in the rapid-proton capture process and their properties
are crucial in defining the flow followed and its termination close to $^{100}$Sn. }

%=================================================================================
% Main author: MP
%\input{Introduction}
%=================================================================================
\section{\textit{Introduction}}
\label{sec:Intro}

An atomic nucleus - a quantum object composed of $Z$ protons (atomic
number) and $N$ neutrons - is held together by strong nuclear forces.
For limited combinations of the $Z$ and $N$ numbers this object
is stable or very long-lived - in nature there are less than 300 such
nuclei. In laboratories, however, more than 3000 unstable, radioactive
nuclei were synthesized so far, and this number is still growing due
to constant progress of acceleration and detection techniques.
This progress is motivated and reinforced by intense research activities
devoted to study properties of nuclei as far as possible from stability,
at the edges of nuclear binding. With increasing imbalance between $Z$ and $N$
numbers, new phenomena appear and new challenges arise for theories of
nuclear structure and reactions.

In general, the body of the nuclear world, as represented on
the chart of nuclei, has three frontiers. One corresponds to large
mass numbers, $A = Z + N$. The range of superheavy nuclei is not
yet known, but it is ultimately limited by fission. The second
border stretches along most neutron-rich side of the chart. This chapter,
however, focuses on the third frontier represented by the most proton-rich,
and thus most neutron-deficient, nuclei. To be specific the
most neutron deficient isotopes of all elements up to uranium are considered.
This nuclear domain abounds
with phenomena and features which either do not have counterparts at the
neutron-rich side of the chart or differ from them significantly.

One of the most characteristic hallmarks here is the Coulomb interaction.
From its interplay with nuclear forces, the Coulomb barrier emerges,
which hinders the emission of unbound protons. As a result, even beyond
the proton dripline, $\beta$ decay competes with proton (\emph{p}) or two-proton
(2\emph{p}) radioactivity and often dominates.
Since the $\beta$-decay energies are large and separation energies
for charged particles are small, various channels of delayed emission
after $\beta^+$ decay are opened. The wavefunction
of a weakly bound proton cannot extend as far as in the neutron-halo case
due to the Coulomb barrier.
On the other hand, the increased radius of \emph{s}-proton states
leads to energy reduction known as the Thomas-Ehrmann
shift, which does not have an equivalent at the neutron-rich side.
Important source of information on nuclear forces are nuclei along the
$N = Z$ line, where protons occupy the same single-particle states as
neutrons. This line, starting at stability for lightest systems, approaches
the proton dripline with increasing mass number and crosses it around
$^{100}$Sn. This supposedly doubly-magic, $N=Z$ nucleus is an
important testing point for the nuclear shell-model.
Another doubly-magic candidate is $^{48}$Ni,
which is the most neutron-deficient nucleus known, having isospin
projection $T_z = -4$. A lot of experimental efforts are still needed
to study these extremely hard to reach exotic systems.
Investigations of nuclei on both sides of the $N = Z$ line probe the
nuclear mirror symmetry and shed light on the limitations and
advantages of the concept of isospin in nuclei.

In this chapter a portrait of the fascinating territory
of nuclei at the proton dripline is sketched.
First, a global overview of the region will be given with the current
status of the exploration. Then, the main methods
used to produce proton rich nuclei will be briefly presented.
Further, several
phenomena and research topics specific to the proton dripline region will be described in more details.

All information on nuclear properties, where no specific reference is given,
were taken from the \cite{NNDC:2022}.

%=================================================================================
% Main author: MP
%\input{Landscape}
%=================================================================================
\section{\textit{Landscape}}
\label{sec:Land}

The limits of nuclear existence are characterized by nuclear binding
energies. The line separating nuclei bound by nuclear forces
from the unbound ones is called the dripline which is defined by means of
nucleon separation energies. Since here the proton-rich
edge of the chart of nuclei is being discussed, the corresponding dripline is
determined by the one-proton and two-proton separation energies:
\begin{eqnarray}
    S_p(N,Z) = B(N,Z)-B(N,Z-1) \\
    S_{2p}(N,Z) = B(N,Z)-B(N,Z-2).
\end{eqnarray}
$B(N,Z)$ is the binding energy of the nuclide (i.e. the neutral atom) defined by:
\begin{equation}
    M(N,Z) = Z \, M_H + N \, m_n - B(N,Z)/c^2,
\end{equation}
where $M(N,Z)$ is the nuclide mass, the $M_H$ and $m_n$ are masses
of the hydrogen atom and the neutron, respectively, and $c$ is the
speed of light. Both separation
energies decrease when moving from stability towards neutron deficient
side along the line of isotopes with a given atomic number $Z$.
Then, the proton dripline is located between the last isotope
with the positive values of $S_p$ and $S_{2p}$ and the next one
for which one of these separation energies becomes negative.
It happens that for the odd-$Z$ elements, the $S_p$ energy
dictates the position of the dripline, as it becomes negative first.
However, for the even-$Z$, due to the pairing energy between protons,
the $S_p$ for the proton-rich systems is larger than the $S_{2p}$,
thus the latter value settles the dripline location.

\begin{figure}[tb]
\begin{center}
\includegraphics[width = 0.8 \columnwidth]{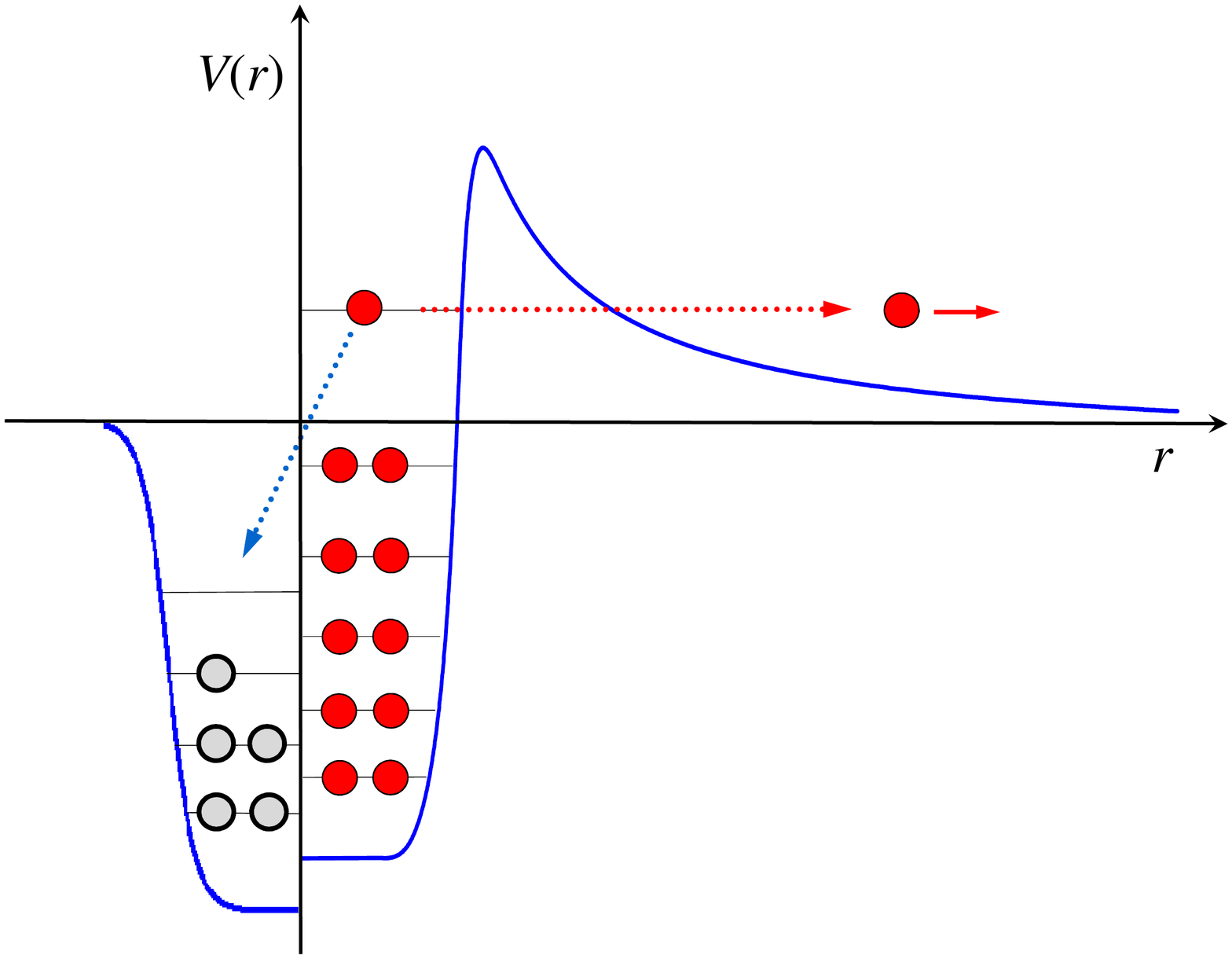}
\caption{Schematic representation of a proton-unbound nucleus. The blue solid line
illustrates the radial part of nuclear potential as a function of the distance
from the nuclear center ($r$). The potential felt by protons, represented by red
circles, is shown on the right side. The sum of the repulsive, long range Coulomb
interaction and the attractive, short range nuclear force creates the Coulomb barrier.
The potential felt by neutrons (gray circles) is shown on the left. The dotted
lines indicate alternative decay modes: \emph{p} emission and $\beta^+$ decay. }
\label{fig:Land_Potential}
\end{center}
\end{figure}

Nuclei beyond the proton dripline are proton-unbound, so they can
spontaneously emit protons from the ground state, giving rise to
one-proton or two-proton radioactivity.
The nuclear potential for a proton-unbound nucleus
is schematically presented in Figure~\ref{fig:Land_Potential}.
Although the last proton is unbound, it is confined by the
Coulomb barrier which prevents it from rapid emission.
The tunneling probability through this potential barrier
is extremely sensitive to the height and the thickness of the barrier,
thus, to the decay energy $Q_{p} = - S_{p}$.
Only when this probability is large enough, the proton emission
can win the competition with $\beta^+$ decay. The same argument
applies to two-proton emission for even-$Z$ nuclei. It follows that
while the observation of proton (or two-proton) emission proves the nucleus to
be located beyond the proton dripline, the position of this line,
except for the lightest nuclei, cannot be determined from decay
data alone. The exact position of the dripline can be deduced
only from precise mass measurements of the nuclei.

For completeness, in the past also other definitions
of the dripline were used, which were based on the half-life criterion
or on the dominance of nucleon emission over $\beta$ decay.
The definition given above, based on separation energies,
is presently most widely accepted and will be used throughout
this article.

\begin{figure}[tb]
\begin{center}
\includegraphics[width = \columnwidth]{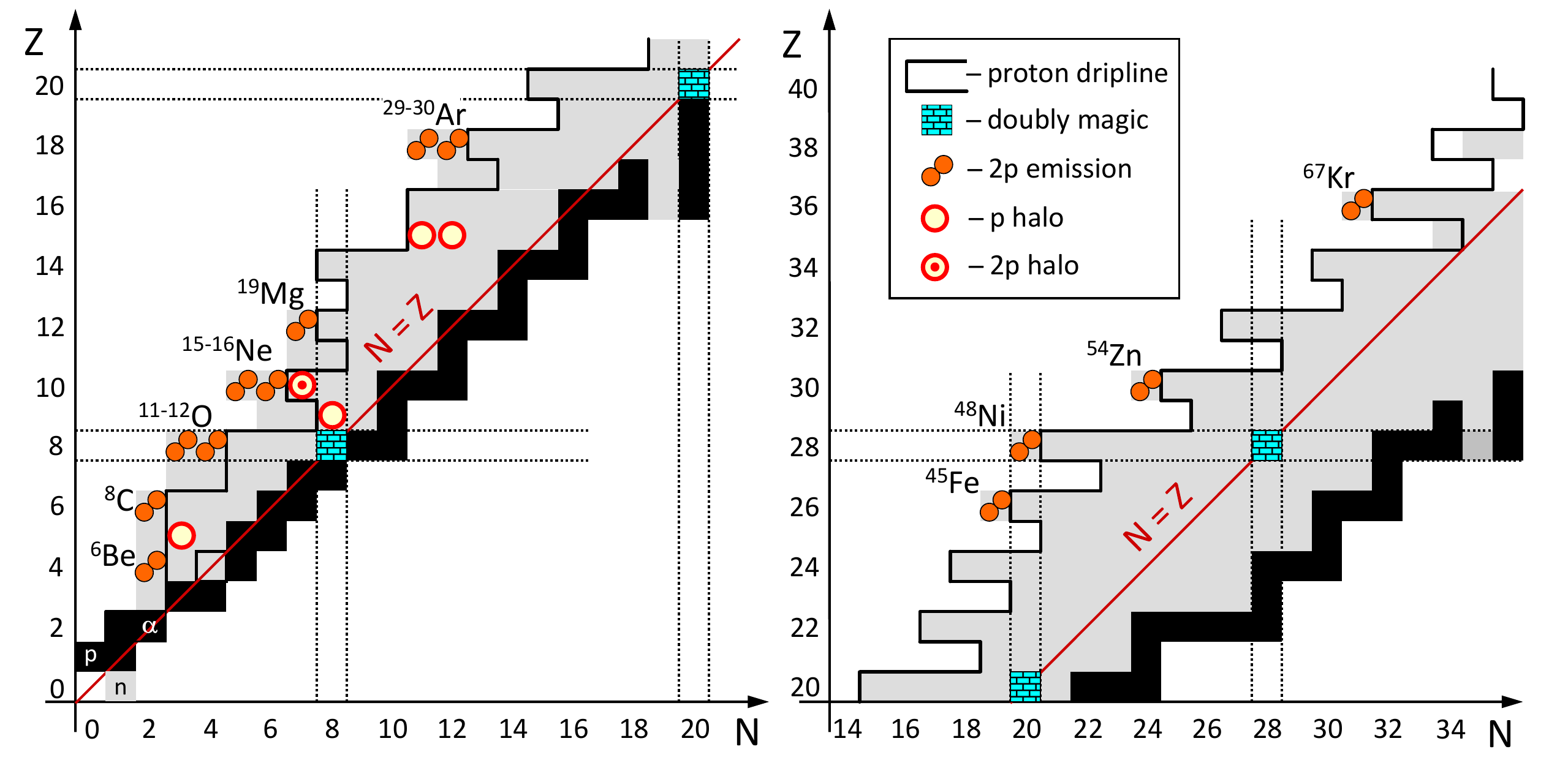}
\caption{Part of the chart of nuclei showing proton rich isotopes
of elements below zirconium ($Z=40$). Labels refer to nuclei
decaying by ground-state 2\emph{p} emission.
The black squares mark the stable nuclei.
Gray area denotes nuclei reached by experiments. The solid line represent the
proton dripline prediction by \cite{Neufcourt:2020, Neufcourt:2020b}.
Dotted lines show the position of magic numbers.
}
\label{fig:Land_Chart_A}
\end{center}
\end{figure}

\begin{figure}[h]
\begin{center}
\includegraphics[width = \columnwidth]{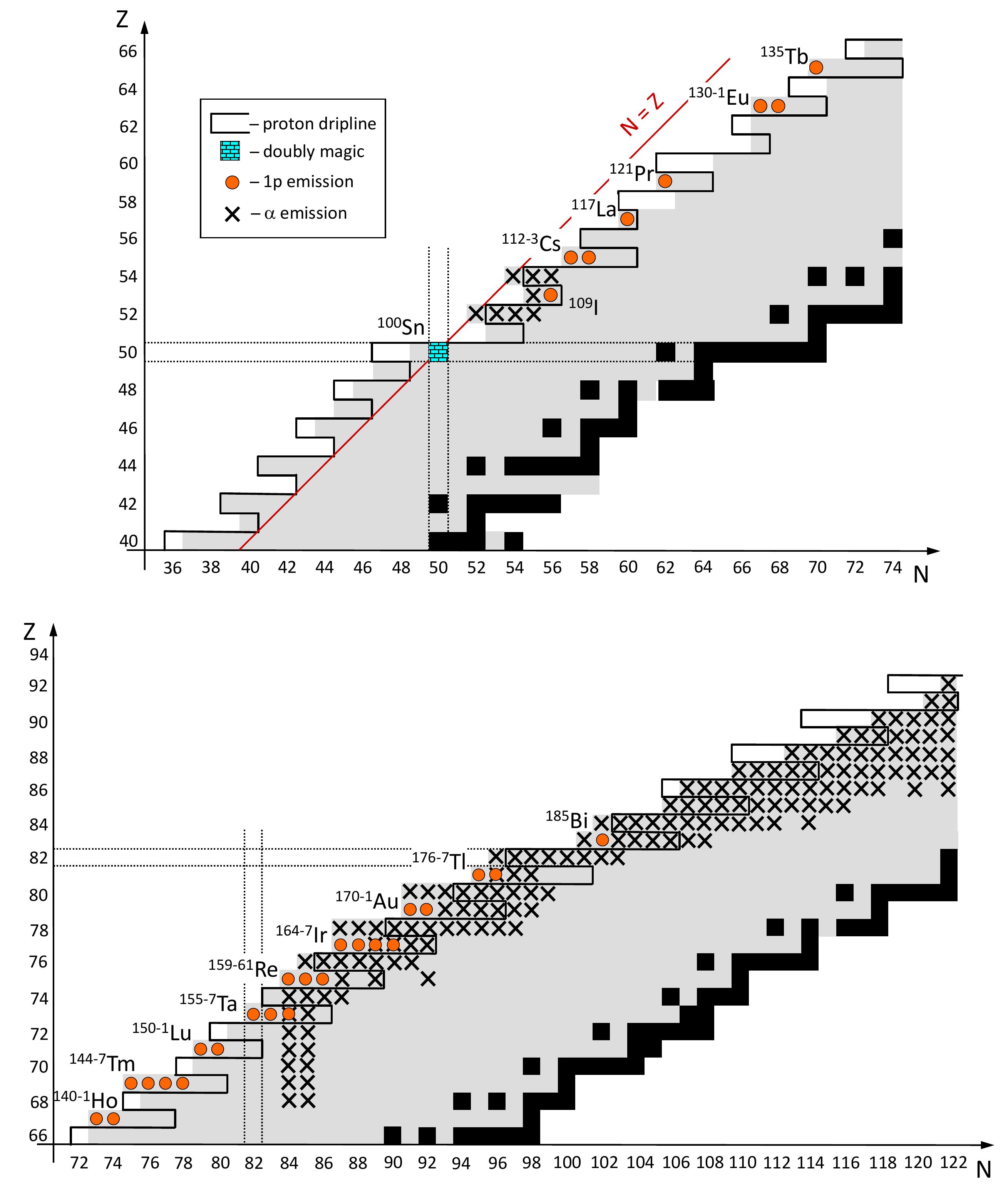}
\caption{Parts of the chart of nuclei showing proton rich isotopes
of elements between zirconium ($Z=40$) and uranium ($Z=92$).
Labels indicate the position of $^{100}$Sn and nuclei decaying
by \emph{p} emission. Crosses mark nuclei for which the
$\alpha$ emission is the dominating decay mode.
Otherwise the same as in Figure~\ref{fig:Land_Chart_A}.}
\label{fig:Land_Chart_B}
\end{center}
\end{figure}

Theoretical modelling of the proton dripline is one of important
probes for understanding of nuclear forces. A recent global prediction,
obtained with an advanced modern method, is shown in
Figures \ref{fig:Land_Chart_A} and \ref{fig:Land_Chart_B}.
The solid lines represent the proton dripline according to
\cite{Neufcourt:2020, Neufcourt:2020b} who
applied the state-of-the-art nuclear density functional
framework with several energy-density functionals to predict masses
at the neutron-deficient edge of the nuclear chart. A novel
feature of their approach was a Baeysian Model Averaging (BMA)
analysis which improves the quality of mass predictions by
including the available experimental information through
machine learning techniques. In this way, the ``collective"
prediction of maximized accuracy, rooted in the current
experimental knowledge, was obtained.

Figures \ref{fig:Land_Chart_A} and \ref{fig:Land_Chart_B}
present also features and hallmarks of the proton dripline
region which will be discussed in the following.
The area in gray represents the nuclei which were
observed, thus for which some experimental information
is available. The established \emph{p}- and 2\emph{p}-emitters are
indicated and their locations can be seen to be consistent with the
predicted dripline. In contrast to the
neutron-rich edge of the chart, the experimental reach is very close
to the proton dripline for all elements and in many cases,
mostly for odd-$Z$, it goes beyond. For example,
the proton-radioactive, most neutron-deficient isotope of
thallium known to date, $^{176}$Tl, has seven neutrons less than the
last bound isotope $^{183}$Tl! Similarly, the most neutron
deficient isotopes of gold ($^{170}$Au), iridium ($^{164}$Ir),
rhenium ($^{159}$Re), and thulium ($^{144}$Tm), all identified
as proton emitters, are located six neutrons beyond the
proton dripline. This is a result of the Coulomb barrier,
which prevents a fast escape of unbound protons.
How far it is possible to go with nuclear spectroscopy of proton-unbound
nuclei depends on experimental limits of production, identification
and detection of the shortest half-lives. The reason why no
proton emitters were observed yet below tin is just because
of the lower Coulomb barrier their expected half-lives are
too short for the current experimental techniques.

In the proton-rich landscape two special landmarks are
$^{48}$Ni and $^{100}$Sn - nuclei with magic numbers
of neutrons and protons, according to the classical
shell model. Whether they indeed are doubly closed-shell
systems is still a matter of studies and attracts a lot
of interest, especially in view of growing evidence for
shell evolution far from stability, as discussed by \cite{Otsuka:2020}.
$^{48}$Ni decays by 2\emph{p} radioactivity (\cite{Pomorski:2011})
and it is believed that detailed studies of this decay mode
may shed light on the structure of this nucleus.
A support for the doubly magic character of $^{100}$Sn
was provided by \cite{Auranen:2018} who observed the
superallowed $\alpha$ decay chain
$^{108}$Xe$\rightarrow ^{104}$Te$\rightarrow ^{100}$Sn,
albeit still with extremely low statistics (two events!).

The appearance of $\alpha$ decay on the chart of nuclei
is a more general indicator of increased binding due to
closed shells. A little island of $\alpha$-decaying
isotopes of tellurium ($Z=52$), iodine ($Z=53$), and
xenon ($Z=54$), above $^{100}$Sn, is the first
manifestation of this effect.
It happens again, more clearly, for neutron-deficient
nuclei with $N\geq84$. In fact, from this point $\alpha$
emission becomes the dominant decay mode along the proton
dripline, only competing with proton radioactivity
for the most exotic odd-$Z$ nuclei.

The third crossing of magic lines, seen in
Figure~\ref{fig:Land_Chart_B} (lower part) occurs at $^{164}$Pb,
which has 15 neutrons less than the expected last bound
lead isotope. Unfortunately, this is much too far
for any reasonable chance to consider it as a subject
of nuclear physics studies.

%=================================================================================
% Main author: CM
%\input{Experimental}
%=================================================================================
\section{\textit{Methods of production}}
\label{sec:Exp}

The study of exotic nuclei at the proton dripline presents several challenges for the
experimenter. It is intrinsic in the {\it exotic} term that such nuclei are likely short-lived
and have very low production rates. Their production cross-section can be as low as a few
femtobarns with rates of the order of single ions/day (\cite{Kubiela:2021}).
In the laboratory, proton-rich isotopes can be produced by fusion-evaporation, fragmentation
or spallation reactions, each of which has its optimum range of applications.

Heavy-ion fusion-evaporation reactions have been over the last decades a very effective
production method for proton-dripline nuclei. Projectiles with energies close to the Coulomb
barrier, typically between 2 and 6~MeV/nucleon, undergo complete fusion with a target nucleus,
creating an excited compound nucleus (CN). Light particles (neutrons, protons, $\alpha$
particles,...) are evaporated from the CN and a residual nucleus is formed. Note that the CN
is more neutron deficient than both the projectile
and the target nuclei, because the $\beta$ stability line turns towards neutron-rich
side with increasing mass. In addition, the evaporation of neutrons is favored over charged particles, which in turn favours the production of ions
more proton-rich than the CN itself. Typical beam intensities amount to several tens of
particle nanoamperes (pnA), while target thicknesses to only a few hundreds of $\mu$g/cm$^2$
up to mg/cm$^2$ to allow for the reaction products to leave the target.

Projectile-fragmentation reactions at mid-to-relativistic energies are an alternative approach
to fusion-evaporation for the production of exotic nuclei, in particular when a suitable
projectile-target combination is not available. It can be pictured as a two-step process:
the first step is the collision (abrasion), which removes nucleons from the overlap region.
The remaining nucleons (or ``spectators'') constitute the highly excited ``pre-fragment'',
which in the second phase of the process (ablasion) de-excites via evaporation of particles
(here again, the evaporation of neutrons is favored)
and $\gamma$-ray emission. The so-called ``Abrasion-Ablation'' Model developed by
\cite{Gaimard:1991} is based on such a view.
Typical beams range from carbon to uranium at energies from 50~MeV/u to 1~GeV/u, while targets
are of the order of g/cm$^2$, 3--4 orders-of-magnitude thicker than those used in
fusion-evaporation reactions, thanks to the larger beam energy.
A key feature of this reaction is that projectile fragments, after escaping the target, move
basically in the same direction as the projectile and with almost the same velocity. This
allows for a fast transport of products to the detection station and facilitates formation of
a radioactive beam.

Intense beams of exotic proton-rich isotopes can be obtained also by spallation reactions. In
this process, an incident light particle (typically proton) at mid-to-high energy hits a
target nucleus, and triggers a series of reactions: nucleons, light charged-particles
(hydrogen and helium isotopes, pions, ...), and heavier (residual) nuclei are emitted, in
addition to products of elastic and inelastic interactions of the projectile with the target.
The residual nucleus is usually rather highly excited and will de-excite by emitting ejectiles
and $\gamma$ radiation. As a result, exotic nuclei are formed among the multitude of reaction
products. The main advantage of this reaction is that beams
of high-energy protons (0.5--1~GeV) with high current, up to 100~$\mu$A are readily
available, and very thick targets can be used. A disadvantage is that the products
are formed at low velocity and have to be extracted from the target.

The common denominator of these production mechanisms is the fact that a wide range of
reaction products are generated, among which are the isotopes of interest. The latter, given
their exotic nature, have production rates that are orders of magnitude lower than those of
the remaining (contaminant) isotopes. Such feature calls for selection of the ions of interest
among the wealth of reaction products in order to study they properties. The selection (and
identification) is achieved with the aid of electromagnetic fields according to the
mass-to-charge ratio ($A/q$), which, for a particle of momentum $p$ and charge $q$, is related
to the magnetic field $B$ via the relation:
\begin{equation}
\frac{A}{q}=\frac{p}{q}=\frac{B\rho}{\beta\gamma}\cdot\frac{e}{c\cdot u},
\end{equation}
where $\rho$ is the radius of particle trajectory, $\beta$ is the particle velocity in units of $c$, $\gamma$ is the Lorentz factor
($1-\beta^2$)$^{-1/2}$, $e$ is the electron charge, and $u$ is the atomic mass unit.

Two complementary methods are typically employed to select and identify exotic ions: in-flight
separation and isotope separation on-line (ISOL).
The in-flight separation is used together with fusion-evaporation and fragmentation
reactions, in which reaction products emerge from the target with substantial
velocity. From this stems one of the main features of this method - its speed.
The time-of-flight of ions through the separator is of the order of 1~$\mu$s
and such short half-lives can be accessed. The method is independent
of the chemical properties of the ion studied. Ions of interest are
selected by means of a combination of magnetic and electrostatic elements, such as dipoles,
quadrupoles, sextupoles, Wien filters, etc. A crucial characteristic
of the in-flight separation is the possibility of the full identification
of ions that arrive to the end of the spectrometer on an ion-by-ion basis.
Products of fusion-evaporation reaction are selected by a recoil separator.
Examples of recoil separators are FMA at Argonne National Laboratory (\cite{Davis:1992}) and
RITU at Jyv\"askyl\"a (\cite{Leino:1995}). Products of fragmentation reaction
are filtered by a fragment separator. Examples of fragment separators are the FRS at GSI
(\cite{Geissel:1992}), A1900 at the National Superconducting Cyclotron Laboratory
(\cite{Morrissey:2003}), LISE at GANIL (\cite{Mueller:1991}) and BigRIPS at RIKEN
(\cite{Sakurai:2008}).

On the other hand, the ISOL method is usually used in conjunction with the spallation
method. The reaction target is coupled to an ion-source. The reaction products are released
from the target into the ion-source, ionised and extracted by means of a few tens of kV
(typical 30-60~kV) acceleration potential. The separation of the ions according to their
mass-to-charge ratio takes place in a uniform magnetic sector field, and the selected ions are
directed to measuring station(s). In contrast to the in-flight separation, ISOL method is
slow. The release times from the ion source range from tens to hundreds of ms, which limits
the range of isotopes accessible. Nevertheless, in favourable cases, the ion-source can allow
for additional selection of the ions of interest from the isobaric contaminants, thanks to its
chemical sensitivity.
Ion sources consist of a very hot cavity in which ionization proceeds through surface
ionization, thermal ionization or interaction of the atom with a plasma. In some cases, better
separation can be achieved by shining an appropriately tuned laser into the hot cavity of the
ion source and exploiting the selectivity of laser resonance ionisation of atoms
(\cite{Fedosseev:2012}). Laser ionization is a powerful separation method to the level that it
may allow to separate the ground state from an isomeric state.
Examples of ISOL facilities are IGISOL at Jyv\"askyl\"a (\cite{Aysto:2001}), ISOLDE at CERN
(\cite{Isolde}), and ISAC-I at TRIUMF (\cite{ISAC-II}).

Modern facilities couple a traditional first stage based on ISOL or in-flight separation with
post-acceleration. Such combination allows to obtain beams of pre-selected radioactive species
at higher energies and excellent ion-optical properties.
This opens possibilities for a broader range of studies with good quality radioactive beams.
Facilities of this type are, e.g. HIE-ISOLDE at CERN (\cite{HIE-Isolde}) and ISAC-II at TRIUMF
(\cite{ISAC-II}).

Production, separation, and identification methods of exotic ions are discussed in greater
detail elsewhere in this Handbook.

%=================================================================================
% Main author: CM
%\input{Decays}
%=================================================================================
\section{Decays of proton-rich nuclei}
\label{sec:decays}

Along the path on the isobaric chain from stability towards and beyond the proton
dripline, nuclear properties evolve and different phenomena appear. Nuclei west of the
$\beta$-stability line are unstable against $\beta^+/EC$ decay, till the dripline is overcome,
where other phenomena like proton and two-proton radioactivity manifest themselves. Along the
path from stability to the dripline,
once the $N=Z$ line is crossed, the Isobaric Analogue State (IAS) in the daughter
nucleus falls within the $Q_{EC}$ window. Both, the initial state and its IAS belong to the
same isospin multiplet and thus have the same structure (as far as nuclear forces are
concerned), so the Fermi transitions, which can occur only between IASs, become observable in
$\beta$ decay. A way to characterize $\beta$ decay is to determine the so-called comparative
half-life or $ft$-value, where $f$ is the phase space integral that depends on the decay energy and
$Z$  (\cite{Wilkinson:1974}) and $t$ is the partial half-life for a given transition. The
$ft$-value is related to the Fermi- and Gamow-Teller reduced nuclear matrix elements squared
$B(F)$ and $B(GT)$ via the relation:
\begin{equation}
ft=\frac{K}{B(F)+(g_A/g_V)^2 B(GT)},
\end{equation}
where
\begin{equation*}
  K = \ln 2 \frac{2\pi^3\hbar^7}{g_V^2 m_e^5c^4}=6144(4)~\rm{s} \, ,
\end{equation*}
and g$_V$ and g$_A$ are the nuclear vector and axial-vector weak coupling constants,
respectively (\cite{Hardy:2020}). Determination of the $ft$ value, which depends on measured
quantities, allows to extract the transition matrix elements to be compared with theoretical
predictions, shedding light on the structure of the states involved. If the matrix elements
can be calculated with sufficient precision, then the coupling constants can be calculated
from measured quantities. A very important application of this idea is the high precision
measurement of the $g_V$ from $0^+ \rightarrow 0^+$ super-allowed Fermi transitions.
It is used to test the Standard Model of fundamental interactions by checking the
conserved-vector-current (CVC) hypothesis of weak interactions
and the unitarity of the Cabibbo-Kobayashi-Maskawa (CKM) matrix.
A comprehensive survey of this research field is provided by \cite{Hardy:2020}.

\begin{figure}[htbp]
\begin{center}
\includegraphics[width=0.8\linewidth]{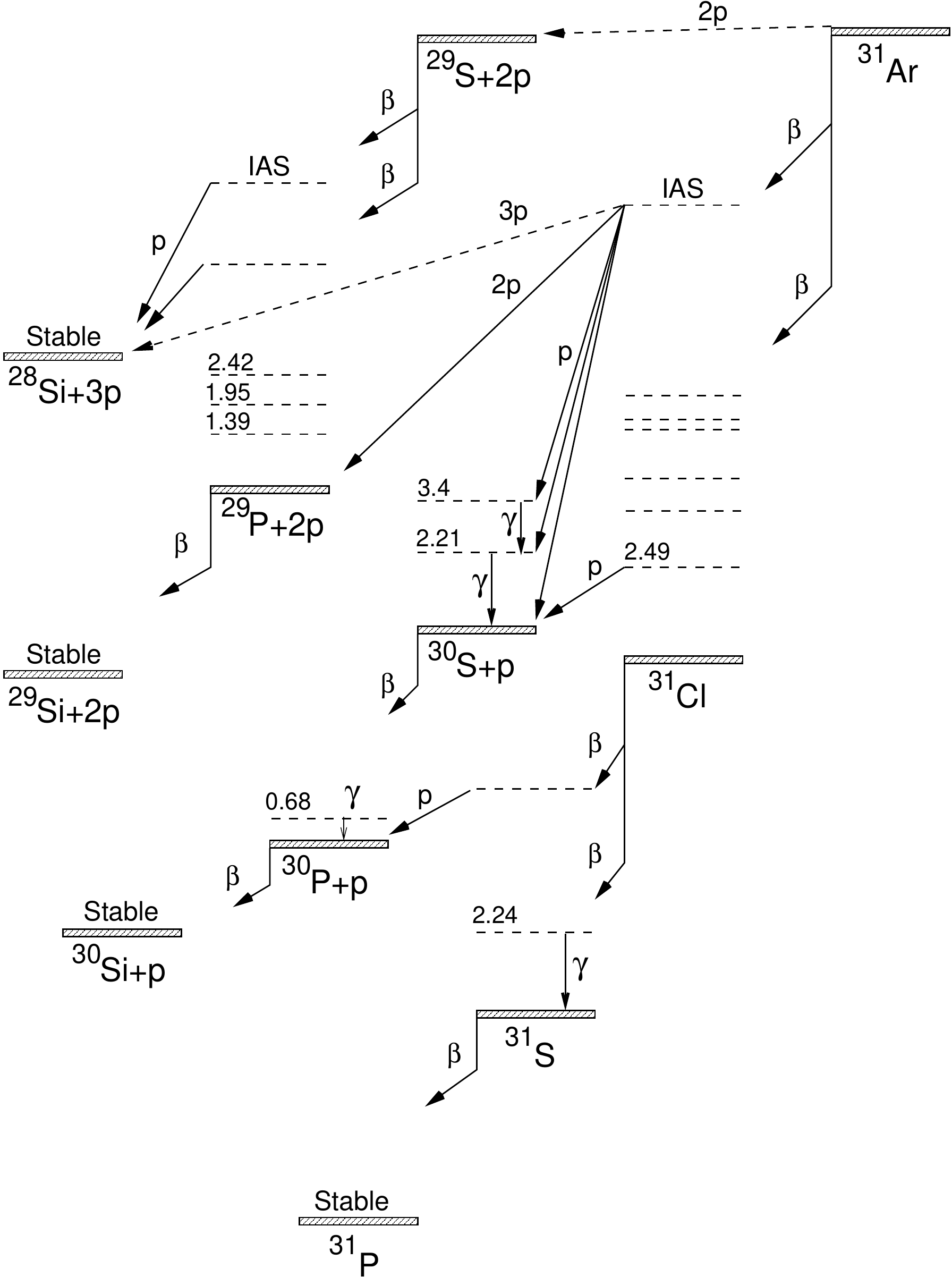}
\caption{Schematic representation of the possible decay paths for $^{31}$Ar, highlighting the
variety of decay channels of an exotic nucleus, as well as the richness of information that
can be acquired by investigating such decay. Note that $\beta 3p$ decay channel of
$^{31}$Ar was reported by \cite{Koldste:2014a} and by \cite{Lis:2015}.
Figure from \cite{Jonson:2001}, Copyright (2001), with permission from Elsevier.}
\label{fig:Decay_31Ar}
\end{center}
\end{figure}

An important characteristic of the path leading from the stability valley to the proton
dripline is the increasing of $Q$-values for $\beta$ decay and simultaneous drop in the
proton(s) separation energies, to the point where the latter will become negative at the
dripline, as introduced in Section~\ref{sec:Land}, see Figures~\ref{fig:Land_Chart_A} and
\ref{fig:Land_Chart_B}.
Such properties open the way to a ``zoo'' of exotic decay modes, ranging from $\beta$ decay
followed by emission of particles (delayed emission) to one- and two-proton radioactivities.
Which $\beta$-delayed emission channels are opened depends on the energy of the daughter state
fed in the $\beta$ decay and on particle separation energies -- in principle any combination
of particles which is allowed energetically can be emitted. So far, for proton rich nuclei,
delayed emission of one, two, and three protons ($\beta p$, $\beta 2p$, $\beta 3p$), as well
as an $\alpha$ particle ($\beta \alpha$) and an $\alpha$ particle with a proton $\beta \alpha
p$ were observed.
The variety of decay channels of an exotic nucleus is illustrated in
Figure~\ref{fig:Decay_31Ar}. Investigating decay properties of so exotic nuclear systems can
provide a wealth of information on nuclear structure of bound or weakly-bound systems.

In the following each of the main decay modes, characteristic for
nuclei at or close to the proton dripline, is presented and illustrated with selected examples. Separate
attention is given to the doubly-magic $^{100}$Sn and the region around it, where the
combination of having $N=Z$ and dripline proximity generates an island of $\alpha$ and proton
decays.

More exhaustive and detailed discussion of exotic decays can be found in review papers
by \cite{Blank:2008}, \cite{Blank:2008c}, and \cite{Pfutzner:2012}.
A recent review of $\beta$-delayed proton and $\alpha$ emission can be found in
\cite{Batch:2020} and the decay mechanisms are also extensively discussed in other parts of
this book.

%%%%%%%%%%%%%%%%%%%%%%%%%%%%%%%%%%%%%%%%%%%%%%%%%%%
\paragraph{\textit{$\beta$-delayed charged particle emission}}

Over a century ago, \cite{Rutherford:1916} observed for the first time $\beta$-delayed
particle emission, $\beta\alpha$ from $^{212}$Bi, although correct interpretation of their
findings came later by \cite{Gamow:1930}. A few decades later, $\beta p$ and $\beta 2p$ decays
were also finally discovered (\cite{Barton:1963,Cable:1983}). All in all, since the discovery
of $\beta$-delayed particle decay, about 200 $\beta p$ emitters are known, some of which
present branching ratios also for $\beta 2p$, $\beta 3p$, $\beta \alpha$ and $\beta p
\alpha$/$\beta \alpha p$ (\cite{Batch:2020}).

Since $\beta p$ stems from population in $\beta^+$/EC decay of excited states that are
particle-unbound, its investigation offers a tool for probing the $\beta$ strength
distribution to particle-unbound states. Decays of nuclei close to the proton dripline, which
have $N\le Z$, have the IAS in the daughter nucleus located well above the particle(s)
separation energies. For these nuclei, Fermi decay to the IAS will have a major contribution
to the decay strength and the position of the IAS makes it open to emission of particles. For
example most of the $\beta 2p$ emission cases known to date proceed in fact through the IAS,
but contribution of Gamow-Teller transitions to higher excited states cannot not be
neglected.
The power of such investigations is well illustrated by the case of $^{31}$Ar, one of the most
studied $\beta$-delayed (multi-) charged-particle emitters, having a large decay energy
$Q_{EC}$=18.3(2)~MeV, see Figure \ref{fig:Decay_31Ar}.

A detailed study by \cite{Fynbo:2000} reported on the measurement of proton energies and
angular correlations in the $\beta 2p$ decay of $^{31}$Ar, showing that the decay mechanism is
dominantly sequential, i.e. the interaction between the proton and the recoil nucleus
determines the decay mechanism, rather than the proton-proton interaction. \cite{Fynbo:2000}
also identified for the first time $\beta 2p$ channels from levels fed by Gamow-Teller
transitions, with consequent impact on the $\beta$ strength distribution.

Observation of $\beta 3p$ emission channel was much more difficult because
a sizable branching ratio for this decay appears in nuclei which
are hard to reach by experiment. In fact, the first observation of this channel
was achieved with a very sensitive and efficient detector, developed mainly
to study 2\emph{p} radioactivity. The TPC-type detector with optical readout
(Warsaw OTPC) was used in a study of 2\emph{p} decay of $^{45}$Fe, as discussed
 a bit later. In the same experiment, however, the $\beta$ decay
branches of $^{45}$Fe were measured and among them the first evidence for
the $\beta 3p$ decay was found with a surprisingly large branching ratio
of 11(4)\% (\cite{Miernik:2007c}). Moreover, ions of $^{43}$Cr
were implanted into the detector as contaminants, and for them
the $\beta 3p$ decay was observed as well with the branching ratio
of $8(3) \times 10^{-4}$ (\cite{Pomorski:2011b}).

\begin{figure}[tb]
\begin{center}
\includegraphics[width = 1.0 \columnwidth]{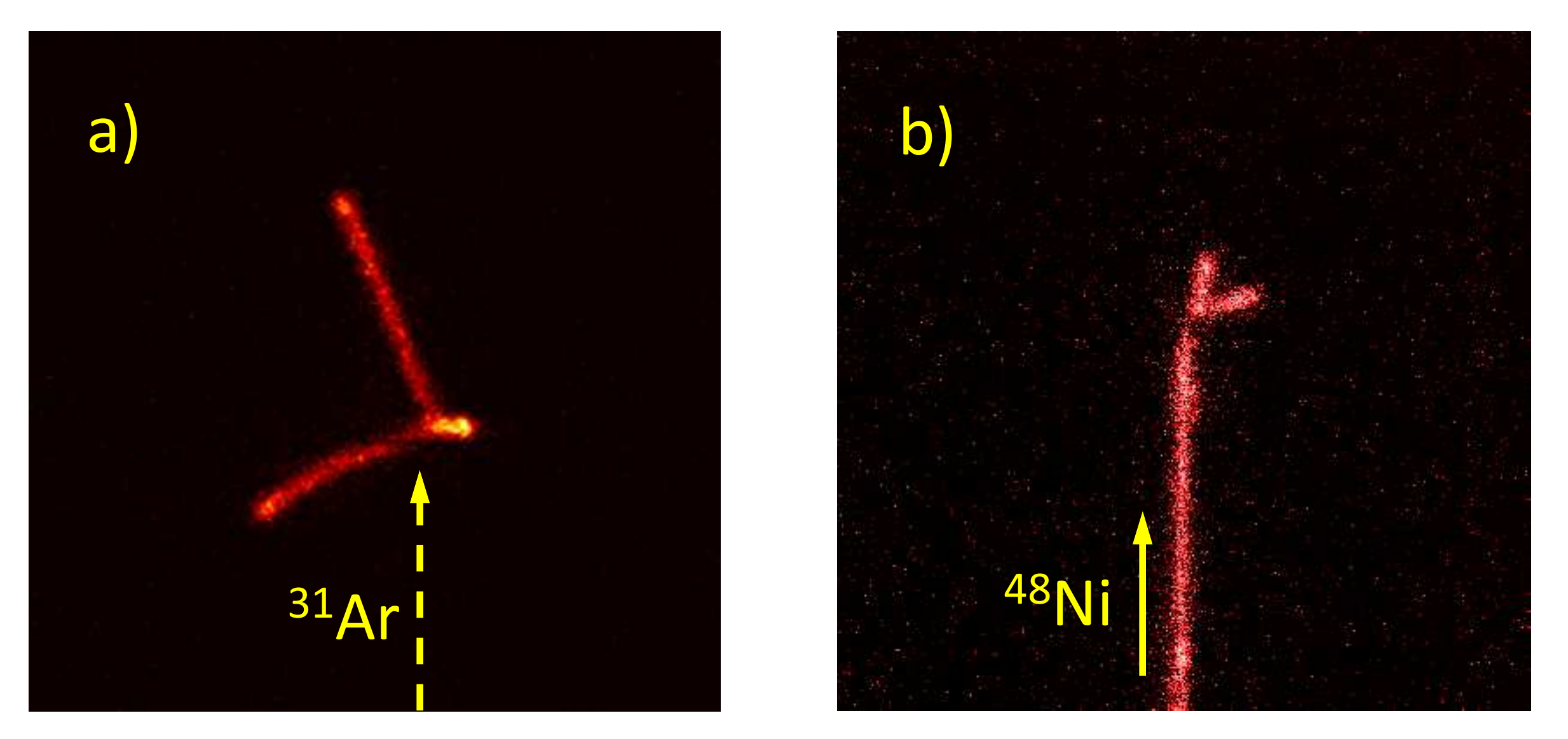}
\caption{Example CCD images recorded by the Warsaw OTPC detector;
a) An event of $\beta$-delayed three-proton emission from $^{31}$Ar. The track of
incoming $^{31}$Ar ion is not seen, as the CCD exposure was started after
the ion implantation. b) Two-proton radioactivity of $^{48}$Ni. Here the
implanted ion entered during the exposure and its track is visible. }
\label{fig:Decay_CCD_images}
\end{center}
\end{figure}

$^{31}$Ar was considered as a candidate for $\beta 3p$ emission
and only after several attempts (\cite{Bazin:1992,Fynbo:1999})
this decay mode was finally identified, again with the Warsaw
OTPC detector (\cite{Lis:2015}), see Figure~\ref{fig:Decay_CCD_images}a.
With 13 events observed, the extracted branching ratio was $7(2) \times 10^{-4}$.
This finding was confirmed by \cite{Koldste:2014a,Koldste:2014} with a different
technique, based on array of silicon detectors.
In spite of its low probability, $\beta 3p$ decay contribution to
the total $\beta$ strength is far from negligible, as was shown by \cite{Koldste:2014}.
About half of delayed three-proton decays proceeds through levels above the IAS in
the daughter $^{31}$Cl and makes up for as much as 30\% of the total Gamow-Teller strength
observed. \cite{Koldste:2014a} shed also some light into the decay mechanism, showing that the
three protons are emitted mainly sequentially.
This remains to date the only case of $\beta 3p$
decay studied to such an extent. The case of $^{31}$Ar highlights the usefulness of
complementary studies with different techniques, as well as the importance
of measuring very weak, exotic decay channels to obtain complete information
on the decay and in consequence on the structure of nuclei involved.

As exotic as $\beta3p$, is $\beta\alpha p$/$\beta p \alpha$ decay.
The two descriptions for the latter reflect the fact that in case of a sequential decay,
this process has two different paths: either the $\alpha$ particle is emitted
first and followed by the proton, or vice versa.
To date, this rare decay
mode has been observed only in three isotopes, $^9$C (\cite{Gete:2000}), $^{17}$Ne
(\cite{Chow:2002}) and $^{21}$Mg (\cite{Lund:2015}). Among these, $^9$C is a special case,
since all states populated in $\beta$ decay disintegrate into a proton and two $\alpha$
particles via the p+$^8$Be and $\alpha$+$^5$Li channels.
For $^{17}$Ne, both $\beta\alpha p$ and $\beta p \alpha$ were observed, with a total branching
ratio of 1.6(4)$\cdot$10$^{-4}$, while the total branching ratio for $^{21}$Mg was found to be
1.6(3)$\cdot$10$^{-4}$.
It is interesting to notice that the known $\beta\alpha p$/$\beta p \alpha$ proceed through
$\alpha$-conjugate nuclei, $^{8}$Be, $^{16}$O and $^{20}$Ne. In their work, \cite{Lund:2015}
attributed the appearance of these rare decay modes to the variation of decay energy caused by
odd-even effects, rather than clustering, as suggested by the path through $\alpha$-conjugate
nuclei.

%%%%%%%%%%%%%%%%%%%%%%%%%%%%%%%%%%%%%%%%%%%%%%%%%%%
\paragraph{\textit{Proton decay}}

Proton radioactivity was discovered over 50 years ago by \cite{Jackson:1970} with the
observation of protons emitted from an isomeric state of $^{53}$Co and about a decade later,
ground-state proton emission was discovered in $^{151}$Lu and $^{147}$Tm by
\cite{Hofmann:1982} and \cite{Klepper:1982}, respectively. To date, more than 40 proton
emitters are known (including decays from isomeric states), as shown in
Figure~\ref{fig:Land_Chart_B} and summarised in the most recent study of proton systematics by
\cite{Delion:2021}. The latter work shows that there is a linear dependence of the logarithm
of the centrifugal-barrier-corrected decay-width from the Coulomb probability multiplied by
the proton-formation probability, in a similar way to the Geiger-Nuttall law for $\alpha$
decay.

Proton radioactivity originates from the tunnelling of the proton through the barrier (see
Figure~\ref{fig:Land_Potential}) and its probability depends strongly on the barrier height
and width seen by the proton, hence on its energy and angular momentum $\ell$. Moreover, it is
very sensitive also to the wave function of the initial and final states involved in the
decay. A unique insight into the structure of proton-unbound nuclei is therefore obtained by
measuring proton energies, half-lives and branching ratios.
Moreover, the proton energy, corrected for the recoil, gives a direct measurement of the
proton separation energy: this is usually the first (and often the only) method to obtain
experimental information on the binding energy of nuclei beyond the proton dripline.

Details of nuclear structure play a very important role in proton decay and small fractions of
the wave-function sometime have large impact on the decay. Another important aspect is that
the majority of proton emitters are deformed nuclei and cannot be described by a simple
spherical single-particle scenario. More complex wave functions are involved, with several
components, of which only few play a role in proton emission. In particular, proton-decay
probabilities can be smaller than they would under the assumption of a spherical potential.

An interesting example in this context is the decay of the proton emitter $^{145}$Tm, which
presents fine structure in its decay. The observation of fine structure is a powerful tool,
since it provides additional insight into the wave function of the states involved in the
decay. In their work, \cite{Karny:2003} studied the proton radioactivity of $^{145}$Tm and
discovered proton transitions to the ground state and to the first excited $2^+$
state in $^{144}$Er. The branching ratio for the latter was found to be
$(9.6 \pm 1.5)$\%. An analysis of the wave function of $^{145}$Tm with a particle-core
vibration model showed that it is dominated by the
$\pi 1 h_{11/2} \otimes 0^+$ component (56\%), while the fine structure transition is due only
to a small fraction of it, the 3\% $\pi 2 f_{7/2} \otimes 2^+$ component.

%%%%%%%%%%%%%%%%%%%%%%%%%%%%%%%%%%%%%%%%%%%%%%%%%%%
\paragraph{\textit{Two-proton decay}}

When an even-$Z$ nucleus is beyond the proton dripline and its
proton separation energy is still positive or very small, two protons can be
emitted simultaneously while the emission of one proton is energetically
blocked or strongly suppressed. The possibility of such process
of 2\emph{p} radioactivity was first noted by \cite{Zeldovich:1960}
and then examined in more detail by \cite{Goldansky:1960}.
The 2\emph{p} emission was observed quite early for light nuclei,
which were relatively easy to produce. The first cases studied were
$^{6}$Be (\cite{Geesaman:1977}), $^{12}$O, and $^{16}$Ne (\cite{KeKelis:1978,Kryger:1995}).
Because of the low Coulomb barrier in these nuclei they decay
so fast that their ground states are characterised by the width
rather than half-life - for example the decay width of $^{6}$Be
is about 100~keV. Such decays belong to the category of resonant
phenomena and were dubbed \emph{democratic} decays by \cite{Bochkarev:1984}.

The first observation of 2\emph{p} radioactivity from a long-lived nuclear
ground state occurred about 40 years after the first considerations
of Zeldovich and Goldansky. It was achieved for $^{45}$Fe in two experiments
employing projectile fragmentation of a $^{58}$Ni beam
by \cite{Pfutzner:2002} and  by \cite{Giovinazzo:2002}. The key factor
in this breakthrough was the full identification
of single ions arriving to the detection setup, which is one of the
crucial advantages of the high-energy fragmentation method.
In both experiments ions of $^{45}$Fe were implanted into
silicon detectors and the evidence of the 2\emph{p} decay came from
the measured total decay energy ($Q_{2p}\approx 1.2$~MeV)
and the decay time ($T_{1/2}=2.6$~ms). The same technique
was used later by \cite{Blank:2005} to discover 2\emph{p} radioactivity
in $^{54}$Zn ($T_{1/2}=1.6$~ms) and by \cite{Goigoux:2016} to
establish this decay mode in $^{67}$Kr ($T_{1/2}=7.4$~ms).

To detect directly and separately both emitted protons
in order to measure the momentum correlations between them,
which carry the information about the mechanism of the 2\emph{p} decay,
a different detection technique was adopted.
It was based on the gaseous time-projection
chamber (TPC) idea. In a TPC, the charged particles ionize the
gas along their tracks and the freed electrons drift with a
constant velocity in a uniform electric field towards a
amplification region, after which they are detected. One such
detector (\cite{Blank:2010}) was used in the first, direct detection of
two protons emitted in the decay of $^{45}$Fe by \cite{Giovinazzo:2007}
and later in the study of 2\emph{p} decay of $^{54}$Zn by \cite{Ascher:2011},
but only seven events could be fully reconstructed.

In another, novel approach the TPC signal is read out by optical
sensors, including a CCD camera. This optical TPC (Warsaw OTPC)
was used in a detailed study of 2\emph{p} radioactivity of $^{45}$Fe
by \cite{Miernik:2007b}.
About 90 2\emph{p} decay events were
recorded and fully reconstructed in three dimensions. The
determined correlation pattern was found to be in a very
good agreement with predictions of the 3-body model developed
by \cite{Grigorenko:2003}. Two important conclusions followed
from this work. First, the essentially 3-body
character of 2\emph{p} emission was confirmed, which means
that this process cannot be approximated by a sequence of
two-body decays. Second, the correlation pattern seems to
depend on the composition of the wave function of the initial
nucleus, thus offering an insight into its structure.

In another experiment, the Warsaw OTPC was used
by \cite{Pomorski:2011,Pomorski:2014}
to discover the 2\emph{p} radioactivity of
$^{48}$Ni ($T_{1/2}=2.1$~ms). An example CCD image of a
2\emph{p} decay of $^{48}$Ni is shown in Figure~\ref{fig:Decay_CCD_images}b.
Since only four such decay events were recorded, no conclusions
concerning the structure of this, supposedly double-magic,
nucleus could be drawn. It is expected, however, that future
experiments on $^{48}$Ni ($Z=28$) and on $^{54}$Zn ($Z=30$)
will provide data with larger statistics and in connection
with results for $^{45}$Fe ($Z=26$) will shed light on the
$Z=28$ shell closure at the proton dripline.

All 2\emph{p} emitters observed below iron, shown in
Figure~\ref{fig:Land_Chart_A}, have so short half-lives that there
is no time to identify them before they decay. They are
produced in the reaction of a radioactive beam with a secondary
target and they decay essentially at the same place, in-flight.
The 2\emph{p} decay is identified by detection of all the decay
products and by the kinematical reconstruction. An example of
such approach, to study 2\emph{p} emission
from $^{16}$Ne, $^{19}$Mg, $^{29}$Ar, and $^{30}$Ar, is described
by \cite{Mukha:2015,Mukha:2018}.

Using the two-neutron knockout from a $^{13}$O beam followed
by invariant-mass spectroscopy of 2\emph{p} decay products, \cite{Webb:2019a}
observed broad states of $^{11}$O. The measured spectrum was
well fitted with the prediction of the Gamow coupled-channel
model, including the $3/2^-$ ground state and three excited
states. An interesting point is that $^{11}$O is the mirror
to famous two-neutron halo $^{11}$Li. The ground state
configurations of both nuclei were found to be similar.

A spectacular nuclear decay was observed for $^{8}$C
by \cite{Charity:2011}. This very exotic carbon isotope was produced
by a neutron knock-out reaction from a beam of $^{9}$C,
followed by detection of all decay products. A sequence
of two 2\emph{p} decays, through the ground state of $^{6}$Be
was found, thus a 5-body decay!

%%%%%%%%%%%%%%%%%%%%%%%%%%%%%%%%%%%%%%%%%%%%%%%%%%%
\paragraph{\textit{$^{100}$Sn and its neighbourhood} }

A very special place in the chart of nuclei is taken by the doubly-magic $^{100}$Sn, which is
the heaviest self-conjugate nucleus being also bound against particle emission. The next $N=Z$
nuclides in the sequence, $^{104}$Te and $^{108}$Xe are in fact $\alpha$ emitters and
predicted to lie beyond the proton dripline. The fact that $^{100}$Sn sits so close to the
proton dripline and is expected to be doubly magic, impacts not only the nuclear structure of
the region, but also the path followed by the rp-process, see Section \ref{sec:rp-process}.
Evidence of its doubly magic character is provided by theoretical predictions as well as by a
series of properties of its decay and of the decay of its neighbours
(\cite{Faestermann:2013}).
Here the main characteristics of this very special nucleus are summarized.

From a classic shell-model perspective, the $N=Z=50$ $^{100}$Sn has $\pi$g$_{9/2}$ and
$\nu$g$_{9/2}$ orbitals filled up, while the $\pi$g$_{7/2}$ and $\nu$g$_{7/2}$ are empty.
Therefore, in a pure single-particle picture, its Gamow-Teller $\beta$ decay
$\pi$g$_{9/2}\rightarrow\nu$g$_{7/2}$ will feed 1$^+$ states in the daughter $^{100}$In, with
dominant, very strong decay to one 1$^+$ state. The latter is the so-called
\emph{super-allowed} Gamow-Teller decay.

The first identification of $^{100}$Sn was achieved by \cite{Schneider:1995} and
\cite{Lewitowicz:1995}.
\cite{Hinke:2012} studied the decay of $^{100}$Sn in more detail and measured the half-life
and determined the $Q_{EC}$ by means of $\beta$ end-point energy. They found the fastest
$\beta$ decay observed to date, with a record-low value for the $\log(ft)$. In a recent
experiment, the decay of $^{100}$Sn was reinvestigated by \cite{Lubos:2019} with 10 fold more
statistics and more accurate values for half-life and $Q_{EC}$ were established, leading to
the revised values $\log(ft)=2.95(8)$, $B(GT)=4.4^{+0.9}_{-0.7}$, which are still record low
and high, respectively, in the whole chart of nuclei.

A signature of the doubly-magic character of $^{100}$Sn combined with its proximity to the
proton dripline is the appearance of an island of proton and $\alpha$ radioactivity just above
it. Pioneers on this topic were \cite{Macfarlane:1965}, who observed for the first time
$\alpha$ decay in the region and concluded that {\it ``these nuclei represent the first
opportunity to study alpha decay from nuclei where the ``valence`` neutrons and protons are in
the same single-particle level, in this case the 1g$_{7/2}$ level. This may give rise to a
kind of ``super-allowed'' alpha decay resulting in large reduced alpha widths.''}. Alpha decay
studies of nuclei above $^{100}$Sn can indeed not only provide evidence for the super-allowed
$\alpha$ decay and consequently double magicity of $^{100}$Sn, but also a deeper insight into
shell evolution along the tin isotopic chain, when fine structure is observed, in a similar
fashion to what is achieved in proton decay studies.

\begin{figure}[ht]
\begin{center}
\includegraphics[width=0.7\linewidth]{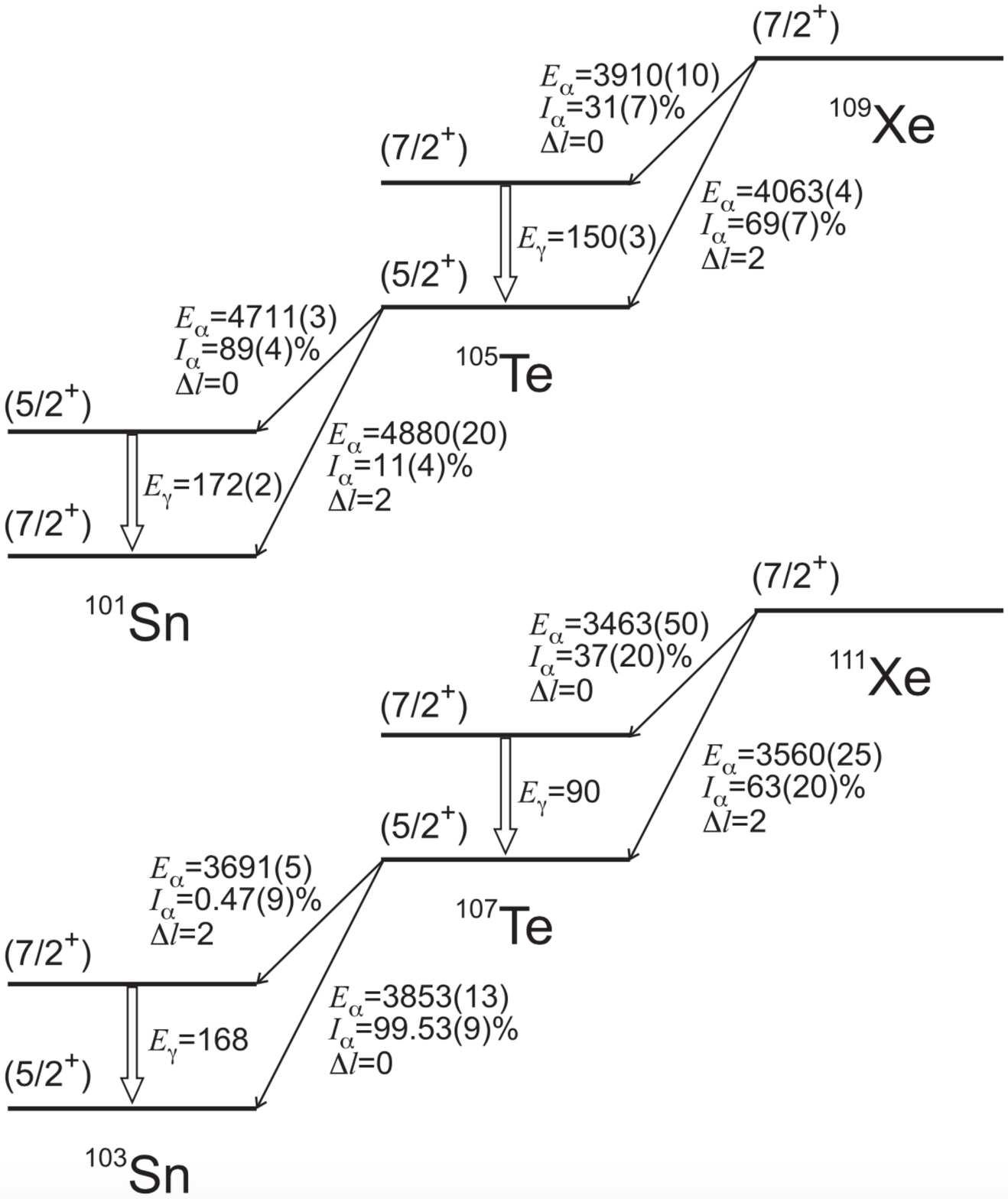}
\caption{$\alpha$-decay chains $^{109}$Xe$\rightarrow$$^{105}$Te$\rightarrow$$^{101}$Sn (Top)
and $^{111}$Xe$\rightarrow$$^{107}$Te$\rightarrow$$^{103}$Sn (bottom). Energies ($E_\alpha$)
in keV, intensities ($I_\alpha$) and angular momentum carried in the decay ($\Delta \ell$) are
shown for each transition, illustrating how spin/parity assignments were obtained. Note the
inversion of levels between $^{103}$Sn and $^{101}$Sn. Figure from \cite{Darby:2010}.
Copyright (2010) by the American Physical Society. }
\label{fig:Decay_109Xe}
\end{center}
\end{figure}

Studies of fine structure in $\alpha$ decay can shed light on the wave  function of the levels
involved in the decay and allow to determine their spin/parity. Of particular relevance is
fine structure in the $\alpha$ decay of odd-A neutron-deficient tellurium isotopes. Several
studies looked at odd-mass xenon-tellurium-tin decay chains as well as directly at
tellurium-to-tin decays
(\cite{Liddick:2006,Darby:2010,Seweryniak:2002,Seweryniak:2006,Seweryniak:2007}). Fine
structure in $\alpha$ decay will populate most likely the lowest-lying excited state in the
daughter nucleus, given the dramatic dependence of the decay probability on the transition
energy. The observation of such phenomenon can therefore allow to determine the position of
the first excited state and combined with the branching ratio lead to the determination of the
spin/parity of the levels involved.
In particular, the decay of $^{105}$Te to the first excited state in $^{101}$Sn allows to
determine the energy separation of the $\nu d_{5/2}$ and $\nu g_{7/2}$ orbitals
(\cite{Seweryniak:2006,Seweryniak:2007}), while fine structure in the decay allows to
determine the order of the orbitals.
\cite{Darby:2010} observed fine structure in both the decays of the
$^{109}$Xe$\rightarrow$$^{105}$Te$\rightarrow$$^{101}$Sn decay chain, leading to levels
positioning with a (7/2$^+$) ground state and a (5/2$^+$) first excited state in $^{101}$Sn,
see Figure \ref{fig:Decay_109Xe}. This observation indicates
a surprising inversion of single-particle orbitals just next to the doubly-magic $^{100}$Sn.

Super-allowed $\alpha$ decay will manifest itself with an $\alpha$ width
\begin{equation}
\delta_{\alpha}^2=h\cdot\frac{\lambda_\alpha}{P},
\end{equation}
much larger than that for the corresponding valence nucleus in the $^{208}$Pb region, the
reference for $\alpha$ decay being $^{212}$Po$\rightarrow$$^{208}$Pb (\cite{Varga:1992}),
which till recently was the only $\alpha$ decay to a doubly magic nucleus known. Here
$\lambda_\alpha$ is the partial decay constant and $P$ the barrier penetration probability.
The $\alpha$ width relative to that of $^{212}$Po, $\delta_{\alpha}^2/\delta^2_{^{212}{\rm
Po}}$, is also called reduced $\alpha$-decay width. (Much) larger-than-one values of the
reduced $\alpha$-decay width have been observed along the decay chains ending in magic
neutron-deficient tin isotopes
$^{114}$Ba$\rightarrow$$^{110}$Xe$\rightarrow$$^{106}$Te$\rightarrow$$^{102}$Sn
(\cite{Mazzocchi:2002,Capponi:2016}) and
$^{109}$Xe$\rightarrow$$^{105}$Te$\rightarrow$$^{101}$Sn (\cite{Liddick:2006,Darby:2010}).
However, the largest value of reduced width (hence $\alpha$ preformation probability) is
expected for the decay of $^{104}$Te leading to $^{100}$Sn and indeed, an enhancement of a
factor of at least 3 was observed in the decay width of $^{104}$Te with respect to $^{212}$Po
(\cite{Auranen:2018}). It should be noted that the latter result stems from a statistics of
only two events and no clear evidence for the $\alpha$ decay of $^{104}$Te could be found in a
separate experiment by \cite{Xiao:2019}, calling for further studies  with improved
statistics. The data available to date on $\alpha$ decay close to $^{100}$Sn have been
recently re-analysed in the context of a superfluid tunnelling model by \cite{Clark:2020},
showing that for nuclei just above $^{100}$Sn the $\alpha$ preformation probability is
significantly larger than for those just above $^{208}$Pb.

Last but not least, another phenomenon that is predicted to happen as a consequence of the
double shell closure at $^{100}$Sn is that of cluster radioactivity, similarly to what happens
in the $^{208}$Pb region (\cite{Bonetti:2007}), with the most promising candidate being
$^{12}$C emission in the decay $^{114}$Ba$\rightarrow$$^{102}$Sn. After a first indication of
this very rare decay mode to exist, a follow up measurement did not confirm the original
results and only an upper limit of 3.4$\cdot$10$^{-5}$ could be inferred for the branching
ratio (\cite{Guglielmetti:1995,Guglielmetti:1997}).
Since the partial half-life for this very rare decay mode is extremely sensitive to the
$Q$-value, its precise determination can help constrain the expected half-life. An
experimental value for $Q_{12C}$ was determined from the observation of the
$^{114}$Ba$\rightarrow$$^{110}$Xe$\rightarrow$$^{106}$Te$\rightarrow$$^{102}$Sn decay chain
(\cite{Mazzocchi:2002,Capponi:2016}) yielding a partial half-life 4-7 orders of magnitude
above the experimentally established lower limit of 1.2$\times$10$^{4}$~s
(\cite{Guglielmetti:1997}). Though, the ideal candidate for observation of $^{12}$C emission
in this region is $^{112}$Ba, which would decay to $^{100}$Sn.
Although $^{112}$Ba production seems to be beyond the possibilities of present-day facilities,
it should be within the reach of the next generation laboratories.

%=================================================================================
% Main author: MP
%\input{Mirror}
%=================================================================================
\section{\textit{Isospin Symmetry}}
\label{sec:Mirror}

One of fundamental principles of nuclear physics is that
to a good approximation the strong nuclear two-body
interactions are charge symmetric and charge independent.
The former means that strong proton-proton
interaction equals the neutron-neutron interaction and the latter
extends this equality also to the proton-neutron interaction.
The consequence of this symmetry is the isospin quantum number,
which should be conserved by strong nuclear forces.
Obviously the isospin symmetry is broken by the Coulomb repulsion
between protons and other differences between the proton and
the neutron, like mass or magnetic moment. The challenge is to
identify and comprehend all the isospin-breaking effects.
In turn, once these phenomena are understood,
the isospin symmetry arguments can be applied to predict properties of
unknown nuclei and anticipate regions of interesting
processes on the nuclear chart. That is why the studies of
isobaric spin (isospin) symmetry become an important research
field in itself.

In this section a few illustrating examples
from the research on isospin symmetry which involve very
proton-rich nuclei are given. The discussion will be limited to pairs
of mirror nuclei, where by mirror the interchange
of $Z$ and $N$ numbers is meant. How asymmetry in mirror
$\beta$ decays can unveil structural differences between proton-rich
nuclei and their reflections on the other side of the $N=Z$ line
will be mentioned in Section~\ref{sec:Halo}, dedicated to proton halos.

\paragraph{\textit{Mirror symmetry}}

According to the charge symmetry, a nucleus and its mirror should
have an identical set of states, having the same spins and parities,
and similar excitation energies. As a result almost all known pairs of
mirror nuclei have the same spin and parity in the ground state.
There are only two known exceptions to this rule. One
is the mirror pair $^{16}_{~9}$F--$^{16}_{~7}$N. The ground state
of $^{16}$F, which is proton unbound, is $0^-$, while that
of $^{16}$N is $2^-$.
There is an excited $0^-$ state in $^{16}$F at 120~keV,
which indicates that there is an inversion of states
between these two nuclei.

Very recently a second case was found for the pair
$^{73}_{38}$Sr -- $^{73}_{35}$Br by \cite{Hoff:2020}.
The ground state of $^{73}$Br is $1/2^-$ and it has an
excited state $5/2^-$ only at 27~keV. In the decay spectroscopy
of the very proton-rich $^{73}$Sr, which is one neutron
away from the dripline, $\beta$ decay to the
IAS in $^{73}$Rb, followed by emission of delayed
protons was measured. Two branches of proton emission were
identified, proceeding to the ground state ($0^+$) and to the first
excited state ($2^+$) in $^{72}$Kr. From the observed intensities
of these two proton transitions, and with help of nuclear
structure calculations using Gamow coupled-channels model
(\cite{Wang:2017}),
the authors concluded that the spin of the IAS must be $5/2^-$,
therefore, the ground state of $^{73}$Sr must be also $5/2^-$.
This is the first breakdown of the mirror symmetry between
ground states of bound nuclei. Although it still needs to be
confirmed, best by a direct measurement of the ground-state
spin of $^{73}$Sr, the inversion of $1/2^-$ and $5/2^-$ states
between $^{73}$Sr and $^{73}$Br was well reproduced
theoretically. \cite{Lenzi:2020} used large-scale shell
model calculations taking into account the Coulomb
interaction and also an isospin-breaking interaction of
nuclear origin, which was introduced phenomenologically.
It was found that the Coulomb interaction plays a dominant
role in the observed inversion, and in fact it would suffice
alone to explain the effect. Nevertheless, room for a small
nuclear contribution was also admitted.

The approach used to explain the state inversion in the
$^{73}$Sr -- $^{73}$Br system was developed originally
for the detailed analysis of energy differences between
isobaric multiplets of high-spin states. It was used
very successfully to study the influence of deformation,
alignment, and intrinsic single-particle configurations
on the mirror energy differences as a function of spin.
\cite{Bentley:2007} summarized this extensive work, focused
mainly on nuclei in the $f_{7/2}$ shell, for which experimental
data for many excited states, up to high spins, were available
and which were still within the scope of
the large-scale shell model. One of the interesting finding
in this program, reported by \cite{Zuker:2002}, was that the
isospin non-conserving interactions of nuclear origin are
at least as important as the Coulomb potential. This type
of studies may bring important new results on the breaking
of isospin symmetry when excited states of heavier nuclei, including
very proton-rich mirror partners, will come within reach
of experiment and large-scale shell-model analysis.

In a few regions of neutron-rich nuclei far from $\beta$
stability changes of nuclear structure have been observed,
exemplified by disappearance of classical shells and emergence
of new ones. A comprehensive summary of these shell evolution
phenomena was given recently by \cite{Otsuka:2020}. One example
is the \emph{island of inversion} around $^{32}$Mg which is related
to the weakening of the $N=20$ shell closure. By mirror symmetry
one could expect corresponding structural changes in very neutron
deficient calcium isotopes ($Z=20$). To probe this idea,
\cite{Doornenball:2007} investigated the mirror pair $^{36}$Ca -- $^{36}$S.
Employing the in-beam $\gamma$ spectroscopy with the $^{37}$Ca beam,
they determined the energy
of the first $2^+$ state in $^{36}$Ca, which was found to be 276~keV lower
than its mirror in $^{36}$S. This large asymmetry was interpreted
as a result of shifts in the single-particle energies and was found
consistent with the expectation that indeed calcium isotopes may
develop another island of inversion for $N<16$.
Shell-model analysis of this case was extended by \cite{Valiente:2018}
to the excited $0^+$ state of intruder type, i.e. formed
by 2 particles - 2 holes excitations. Such state is known in
$^{36}$S at 3346~keV. The calculations reproduced well the observed
difference between $2^+$ states and predicted a much larger asymmetry
for the intruder $0^+$ states - in $^{36}$Ca it should be located
720~keV lower than in $^{36}$S. In fact, it is expected to be the
first excited state in $^{36}$Ca decaying by $E0$ transition to the
ground state. Thus, perspectives of further studies in the dripline
region around $Z=20$ appear attractive, although the mirror of $^{32}$Mg,
$^{32}$Ca, is probably too far beyond the dripline (3 neutrons),
to be reached by experiment.

\paragraph{\textit{Thomas-Ehrman shift }}

\begin{figure}[tb]
\begin{center}
\includegraphics[width = 0.9 \columnwidth]{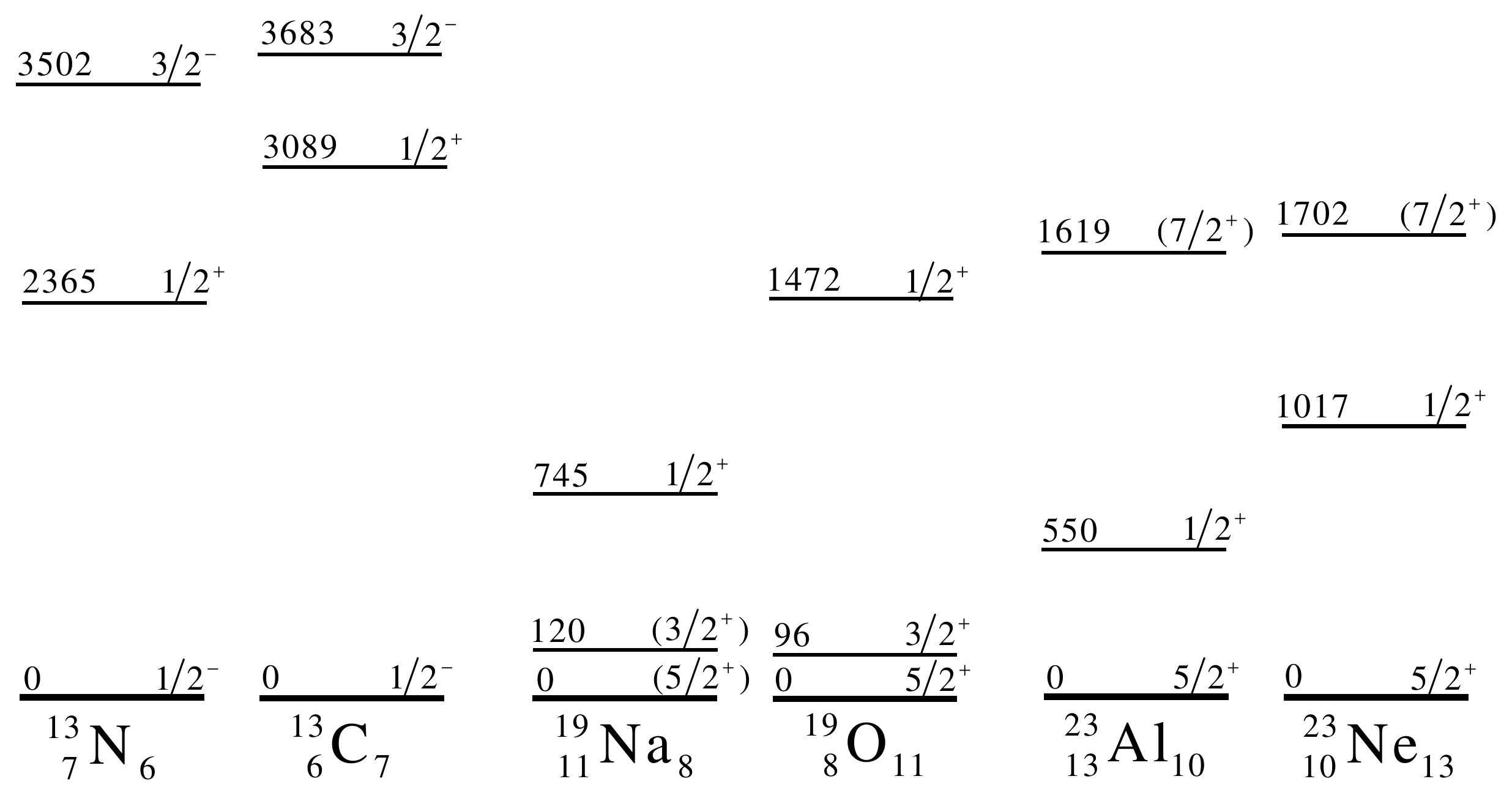}
\caption{Partial level schemes of three pairs of mirror nuclei illustrating
the Thomas-Ehrman shift. Note that the largest energy difference occurs
for $1/2^+$ states - their energy is lower in the proton-rich nucleus.
Energy values are in keV.}
\label{fig:Mirror_TES_examples}
\end{center}
\end{figure}

One reason for an energy shift between mirror states is the increased
radial extent of the weakly bound valence proton in the $s_{1/2}$ orbit.
Levels with significant contribution of this component are expected
to experience smaller Coulomb repulsion and thus to be more bound
(have lower energy) than their neutron-rich mirror partners.
For example, the first excited state in $^{13}$N ($1/2^+$ at 2365~keV)
is 724~keV lower than its mirror in $^{13}$C.
This difference was analysed for the first time by \cite{Ehrman:1951}
and \cite{Thomas:1952}, after whom the effect is named
the \emph{Thomas-Ehrman shift} (TES).

The TES was observed in a few other nuclei where the proton
$s_{1/2}$ orbital plays an important role. Three prominent
examples are shown in Figure \ref{fig:Mirror_TES_examples}.

\begin{figure}[tb]
\begin{center}
\includegraphics[width = 0.8 \columnwidth]{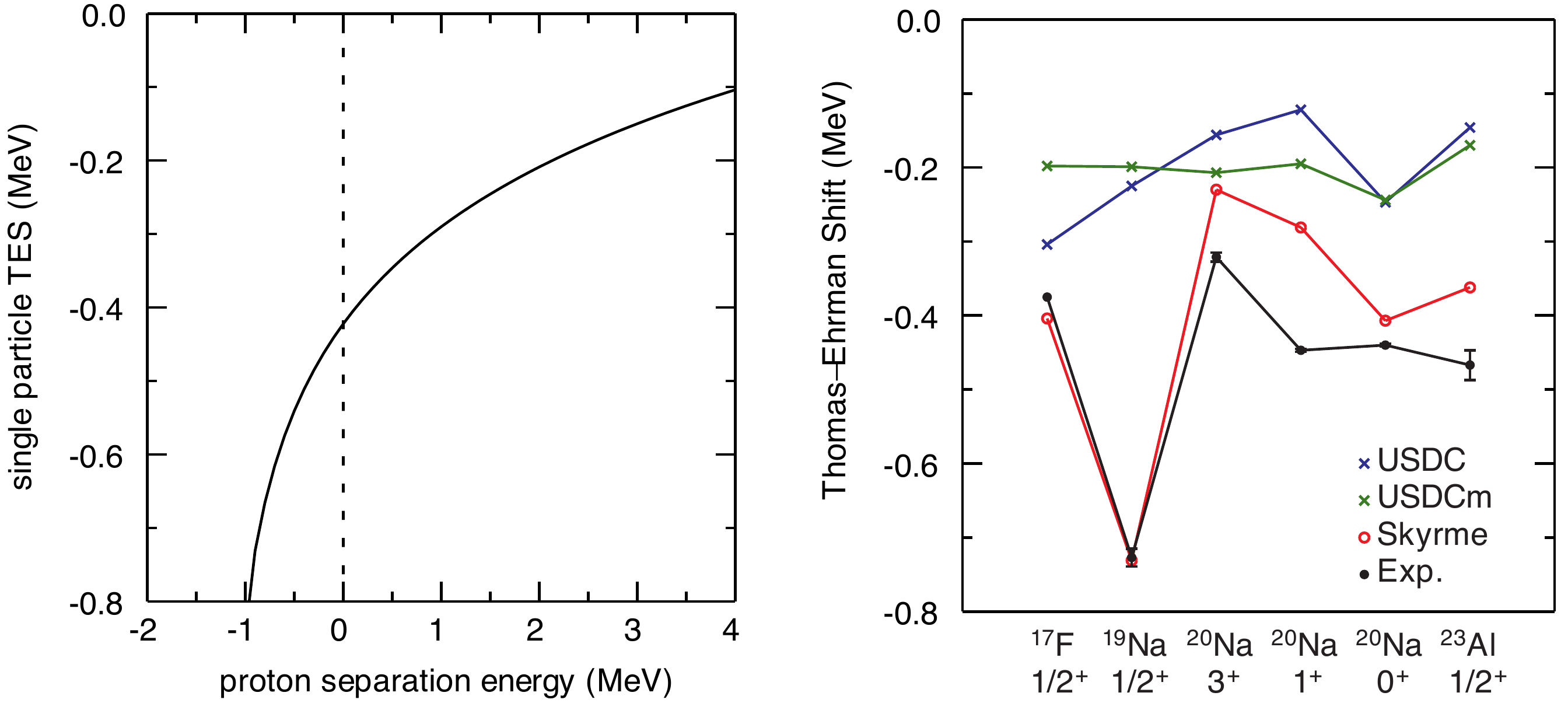}
\caption{Left: Calculated Thomas-Ehrman shift of the single-particle,
valence $s_{1/2}$ proton orbit as a function of proton separation energy.
Right: measured Thomas-Ehrman
shift for the levels indicated (black, full circles), compared with
shell-model predictions; the model including TES is labeled Skyrme
(red open circles). See text for more details. Taken from \cite{Magilligan:2020}, Copyright (2020) by
the American Physical Society. }
\label{fig:Mirror_TES_model}
\end{center}
\end{figure}

Recently a modern shell-model hamiltonian for the $sd$ shell was
developed by \cite{Magilligan:2020}, which incorporates isospin-breaking
interactions. In addition, the authors examined a model of the TES.
First, they calculated how the difference in proton and neutron $2s_{1/2}$
single-particle energies, relative to the difference in $1d_{5/2}$
energies, depend on the proton separation energy. Results for
a $^{16}$O core were found very similar to those with a $^{28}$Si
core, so a single curve, shown in Figure~\ref{fig:Mirror_TES_model} left,
was adopted as a model of the single-particle TES.
This, together with the calculated proton $2s_{1/2}$ spectroscopic
factors, allowed them to determine the energy shift of levels in several
nuclei having $s_{1/2}$ components. The results of this procedure showed
good agreement with experimental values, see Figure~\ref{fig:Mirror_TES_model} right.

Originally, the mechanism of TES was invoked to describe a change of
motion of a single nucleon. The underlying idea, however, can be extended
to groups of nucleons, like clusters. For example $\alpha$-clusters
do appear in the structure of excited states in light nuclei,
especially close to the threshold for $\alpha$ emission. In such case
the nucleus can be viewed as a weakly coupled system of an
$\alpha$-cluster and the residual nucleus. Usually the wave function
of the relative motion of these two parts has a large $s$-wave component
and a difference due to Coulomb interaction should be observed
for the mirror systems. This idea was applied by \cite{Ito:2016}
to the mirror pair $^{10}$C -- $^{10}$Be, which can be described
as two $\alpha$ particles and two nucleons. The calculation of
the $0^+$ states in these nuclei showed indeed significant energy
reduction of states in $^{10}$C which had dominant $s$-wave component.
Similar conclusions were made by \cite{Nakao:2018} for the
$^{18}$Ne -- $^{18}$O pair which were considered in the cluster
model as $\alpha + ^{14}$O and $\alpha + ^{14}$C systems, respectively.
The need to verify these predictions will hopefully motivate
experimental activity in the field of clustering phenomena.

%=================================================================================
% Main author: MP
%\input{Halos}
%=================================================================================
\section{\textit{Proton halos}}
\label{sec:Halo}

The phenomenon of a nucleon halo occurs when, due to a very
low binding energy, one or two nucleons are able to tunnel far
away from the nuclear binding potential. The resulting large
spatial extension is one of the main characteristic features of
a halo state. The neutron halo physics started from the discovery
by \cite{Tanihata:1985} of the large radius of the $^{11}$Li nucleus.
Soon it developed into an important and broad research field
for neutron-rich nuclei and it is extensively
discussed in other parts of this Handbook. Analogous phenomena
are expected also for weakly bound proton-rich nuclei.
Due to the Coulomb barrier, however, the halo effects are
less pronounced and thus more difficult to investigate.
Nevertheless, the one-proton- or two-proton-halos are expected
for low $Z$ nuclei with weakly bound valence protons
in $s$- or $p$- states. Here a few, most
pronounced cases, are presented.

\paragraph{\textit{$^{8}$B}}

The first case established as proton halo, and also the most
intensely studied is $^{8}$B. It has a very low proton separation
energy, $S_p = 136$~keV. It appeared first as a halo candidate
when \cite{Minamisono:1992} measured a large quadrupole moment ($Q$)
of $^{8}$B, in fact twice larger than shell-model predictions.
This result was found to be in agreement with the assumption of
increased charge root mean square radius. Although the
large value of the $Q$ moment is not a sufficient evidence for the
halo structure, as noted by \cite{Nakada:1994}, who showed that
a similar effect can result from E2 core polarization, the case
of $^{8}$B drew the attention to the possible existence of proton halo.

The firm evidence for the extended
spatial distribution of $^{8}$B was provided by \cite{Schwab:1995}
in a measurement of the longitudinal momentum distribution of $^{7}$Be
following the break-up reaction of $^{8}$B beam on different targets.
This method is based on the theoretical prediction that the longitudinal
momentum distribution (i.e. parallel to the beam direction) resulting
from the fragmentation of weakly-bound projectiles is insensitive
to details of break-up interactions and provides a reliable probe
of the internal momentum wavefunction of the system (\cite{Bertulani:1992}).
For a large  spatial extension of the wave function, a narrow momentum
distribution is expected. Indeed, using a beam of $^{8}$B at
about 1.5 GeV/nucleon,
\cite{Schwab:1995} found a narrow distribution of one-proton-removal
product, $^{7}$Be, with a FWHM of $81\pm6$~MeV/c for carbon, aluminum,
and lead targets. This width is smaller by a factor of about 3 from
the prediction of the statistical model of fragmentation
by \cite{Goldhaber:1974} and from the measured values for one-proton
removal from other light projectiles.

Another evidence for the unusually large radius of $^{8}$B was found
soon afterwards.
\cite{Warner:1995} discovered that the total reaction
cross sections, $\sigma_R$, for this nucleus on a silicon target
was larger than for heavier $^{12}$C and $^{14}$N projectiles and
notably larger than from conventional calculations.

\begin{figure}[tb]
\begin{center}
\includegraphics[width = 0.8 \columnwidth]{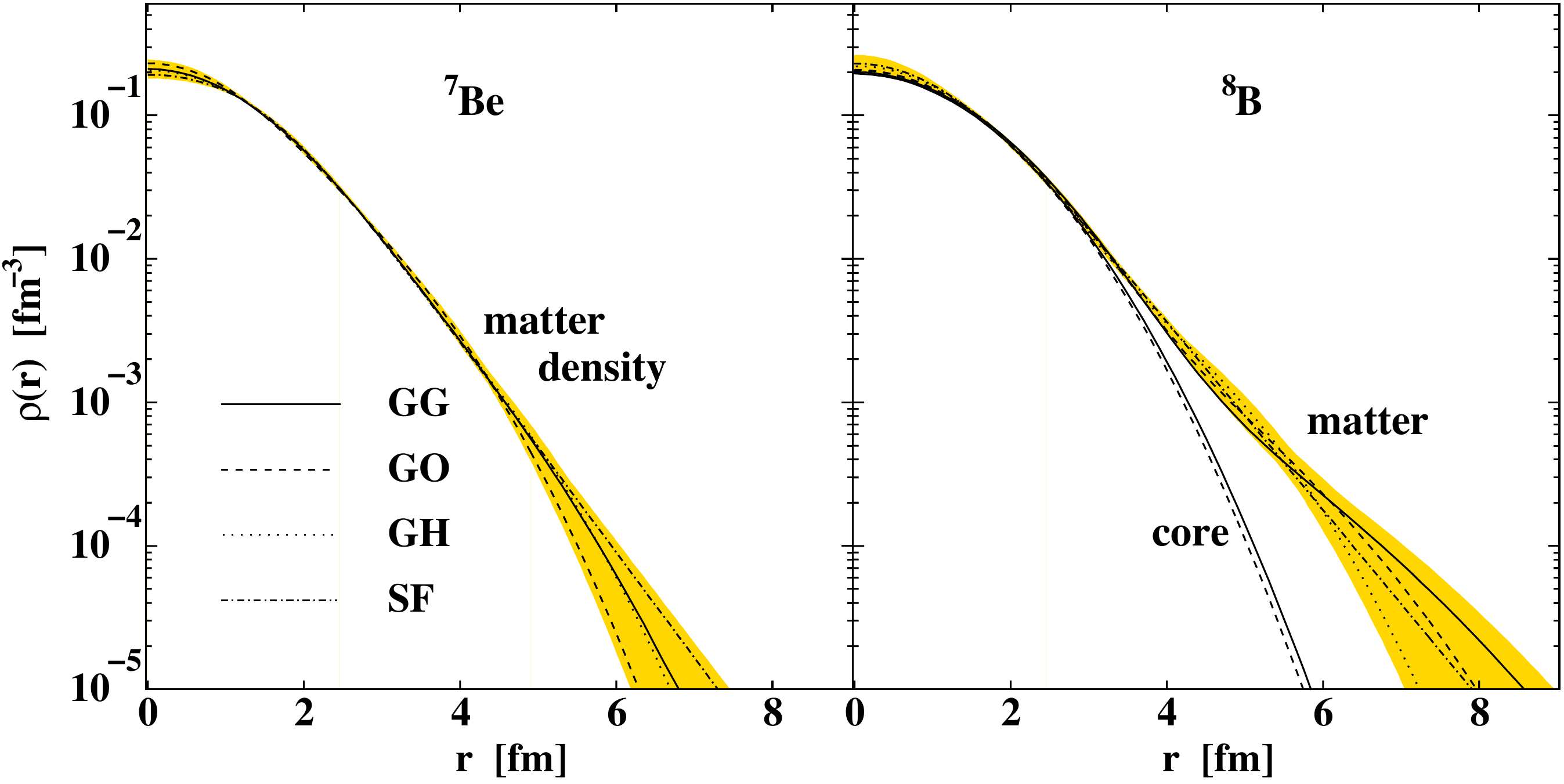}
\caption{Nuclear matter density distributions for $^{7}$Be and $^{8}$B deduced from the
experimental data. The $^{8}$B nucleus was assumed to consist of a $^{7}$Be core and a loosely
bound valence proton. The yellow error-bands represent
envelopes of the density variation within the four parametrizations of density
distributions considered, superimposed by the statistical errors.
Reprinted from \cite{Dobrovolski:2019}, Copyright (2019), with permission from Elsevier. }
\label{fig:Halo_8B}
\end{center}
\end{figure}

Two recent results on $^{8}$B give a flavour of modern, advanced studies of proton halos.
Differential cross sections
for elastic proton scattering on $^{7}$Be and $^{8}$B at an energy 0.7~GeV/nucleon,
in inverse kinematics, were measured by \cite{Dobrovolski:2019}.
Beams of $^{7}$Be and $^{8}$B impinged on an ionization chamber
filled with pure hydrogen at a pressure of 10 bar,
which acted as a target and at the same time as
a detector of the recoiling proton (active target).
The results were analyzed using Glauber
multiple-scattering theory and the matter density distributions
were determined. The applied formalism is described in
detail by \cite{Alkhazov:1978} and the results are shown
in Figure~\ref{fig:Halo_8B}.
By adding one proton to $^{7}$Be, the tail
of the matter distribution changes significantly.

Elastic scattering of $^{8}$B on $^{64}$Zn at much lower energy, close to the
Coulomb barrier, was investigated by \cite{Sparta:2021} using a
post-accelerated $^{8}$B beam. The measured data were interpreted with
help of the continuum-discretised coupled-channels (CDCC) calculations
(\cite{Thompson:1988}). The very interesting finding was that in spite
the extended matter distribution of $^{8}$B, its reaction dynamics
is very different from that of neutron halo nuclei. In particular,
the angular distribution does not show a reduction in the Coulomb-nuclear
interference region, like observed for the neutron-halo of $^{11}$Be.
This different behaviour is interpreted as due to additional Coulomb
halo-core and halo-target interactions. This shows also that the full
understanding of the proton-halo dynamics is far from complete even
for this best studied case.

\paragraph{\textit{$^{17}$F}}

The ground state of $^{17}$F is $5/2^+$, as expected when a proton
is added to the $^{16}$O core and occupies the $1d_{5/2}$ orbital.
The next orbital, $2s_{1/2}$, gives rise to the first excited state
$1/2^+$ at 495~keV. Since the proton separation energy from this state
is only 105~keV, it is a good candidate for the proton halo.
The first hint of that was noticed in the measurement of $\beta$ decay
of $^{17}$Ne by \cite{Borge:1993}. They investigated the
first-forbidden transition into the first excited
state of $^{17}$F and compared this decay channel with the corresponding
mirror transition of $^{17}$N to $^{17}$O. The difference between the two
transitions is quantified by the parameter $\delta_{\beta}$:
\begin{equation}
  \delta_{\beta} = \frac{(ft)^+}{(ft)^-}-1 \, ,
\end{equation}
where $ft^{+/-}$ is the comparative half-life for $\beta^{+/-}$
transitions where one is a mirror of the other.
In case of strict mirror symmetry $\delta_{\beta}$ should be equal to 0,
while \cite{Borge:1993} found $\delta_{\beta} = -0.55(9)$. This large asymmetry
was explained by the substantial extent of the proton $2s_{1/2}$ orbit and the
halo structure of the $^{17}$F state was postulated. However, as noted
by \cite{Ozawa:1998}, who confirmed the value of $\delta_{\beta}$,
this finding could also point to an abnormal structure of the $^{17}$Ne
ground state, which is a halo candidate itself, as discussed later.

The confirmation of the halo in $^{17}$F came from a measurement
of the capture cross section for the reaction $^{16}$O$(p, \gamma)^{17}$F
by \cite{Morlock:1997}. The study was motivated by nuclear astrophysics,
as this proton capture reaction is important for the CNO cycle of
hydrogen burning in second-generation stars. The authors found that
the capture to the first excited $1/2^+$ state in $^{17}$F strongly
increases with the decreasing energy of protons. This could be
explained only by the halo properties of the $1/2^+$ state.
A direct-capture model suggested that at the energy of 100~keV (in the
center of mass), the main contribution to the capture to the $1/2^+$ state
comes from a distance of about 50~fm from the nucleus!

\paragraph{\textit{$^{26}$P and $^{27}$P}}

One of the earliest theoretical predictions of proton-halo candidates
was made by \cite{Brown:1996}. Using the shell model they analysed
the proton-rich nuclei in the $sd$ shell and suggested three halo
candidates: $^{26,27}$P and $^{27}$S. %To the latter case we come
%back a little later, here we deal with isotopes of phosphorus.
The latter case will be mentioned a little later, while in this
section the isotopes of phosphorus are considered.

The phosphorus isotopes ($Z=15$) are the lightest nuclei expected
to have the last proton in the $2s_{1/2}$ orbital in the ground state.
The proton separation energy for $^{26}$P and for $^{27}$P is 140~keV
and 870~keV, respectively. \cite{Navin:1998} measured longitudinal momentum
distribution of projectile residues after single-nucleon knockout
reactions from the beams of phosphorus isotopes.
The widths of these distributions were
found indeed narrow, as expected for halo structures. Furthermore,
the data were consistent with the assumption that the ground states
of $^{26}$P and $^{27}$P are dominated by the $s$ component.
The same observation was actually made also for $^{28}$P having
$S_p = 2052$~keV, which is less favourite for the
halo formation.

\cite{Fang:2001} produced a number of proton-rich nuclei by a fragmentation of
$^{36}$Ar beam at the energy of about 70~MeV/nucleon and
measured for them the total reaction cross section, $\sigma_R$,
on a carbon target. The value for $^{27}$P was found to be
significantly larger than for neighboring nuclei. The density
distribution of $^{27}$P, determined with help of Glauber-model
analysis, revealed the long tail reaching above 12~fm, confirming
the proton halo behaviour. No such evidence, however, was
found for $^{28}$P, and no data for $^{26}$P were obtained.

In turn, more light on structure of $^{26}$P was shed by a thorough study
of its $\beta$ decay by \cite{Perez:2016}.
One of the results was the determination of $\beta$ strength to
various final levels in $^{26}$Si.
In particular, the strength to the first excited $2^+$ state was
compared with the corresponding mirror transition $^{26}$Na$(\beta \gamma) ^{26}$Mg
yielding the large mirror asymmetry $\delta_{\beta} = 0.51(10)$.
The theoretical analysis by \cite{Kaneko:2019} showed that
this finding can be explained by isospin non-conserving forces
related to the loosely-bound $2s_{1/2}$ orbit, thus
supporting the proton halo hypothesis for $^{26}$P.

\paragraph{\textit{$^{17}$Ne}}

$^{17}$Ne is the last bound neon isotope, it has the Borromean character
($^{16}$F is unbound) and has $S_{2p}=933$~keV --
all this makes it a candidate for the two-proton halo.
The first firm confirmation of this proposition came from \cite{Kanungo:2003}.
Using the fragmentation of $^{20}$Ne primary beam at 135~MeV/nucleon,
they produced a beam of $^{17}$Ne,
which was directed on a 0.5 mm thin beryllium target.
The longitudinal momentum distribution for the two-proton
removal was measured, together with the two-proton removal
cross section ($\sigma_{-2p}$) and the total interaction cross
section ($\sigma_I$). The width of the momentum distribution
was found twice narrower than expected from the \cite{Goldhaber:1974} model.
In addition, the only consistent description of all measured
observables could be obtained assuming that both valence
protons occupy the $2s_{1/2}$ orbital. This suggests that
between the $^{17}$N and its mirror $^{17}$Ne the inversion
of $1d_{5/2}$ and $2s_{1/2}$ orbitals takes place, which is an
example of shell evolution.

\begin{figure}[tb]
\begin{center}
\includegraphics[width = 0.8 \columnwidth]{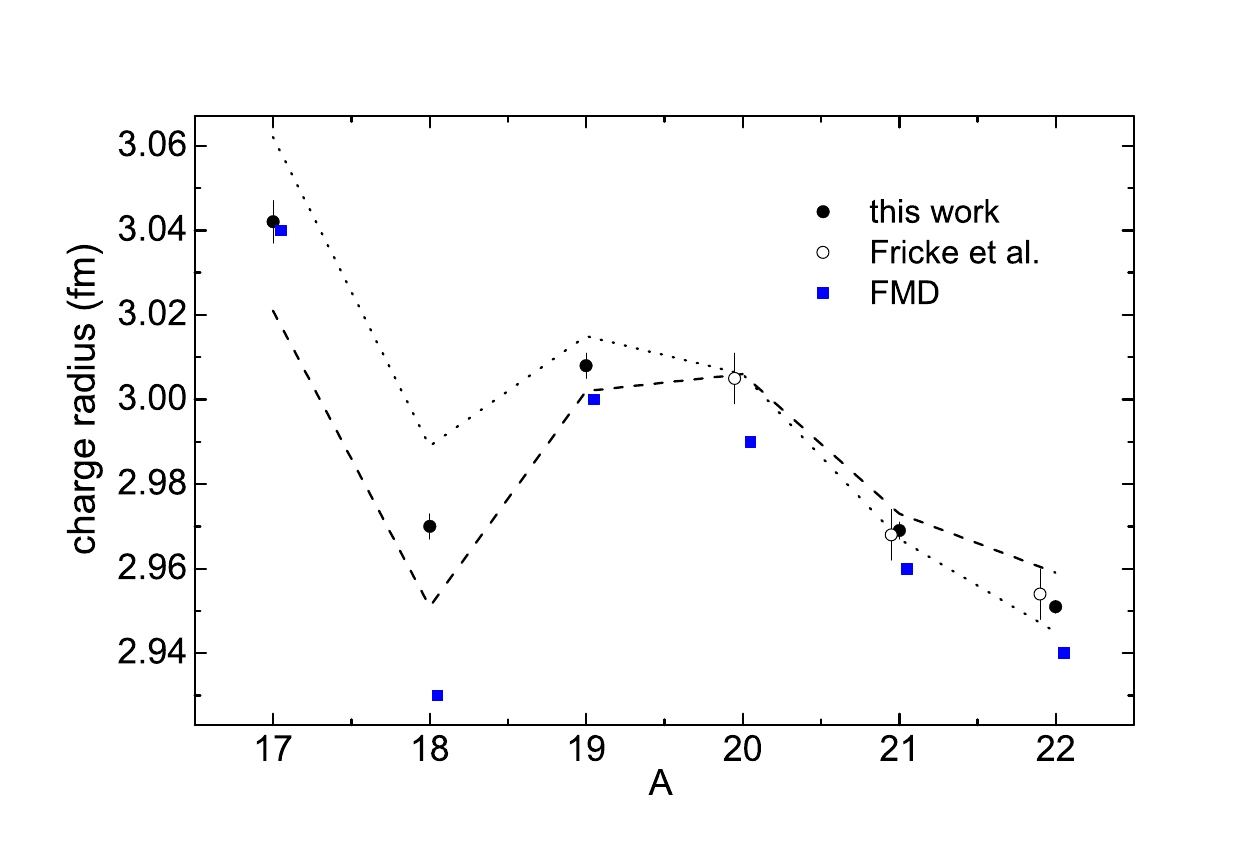}
\caption{The rms charge radii for the neon isotopes with mass number $A$.
Error bars indicate the statistical uncertainties. The systematic error
limits are represented by the dotted line and the dashed line.
From \cite{Geithner:2008}. Copyright (2008) by the American Physical Society. }
\label{fig:Halo_Ne-radii}
\end{center}
\end{figure}

Collinear laser spectroscopy is a modern technique for a very precise
measurements of nuclear ground-state properties. It is based
on the hyperfine interaction between nucleus and atomic electrons.
With help of lasers the tiny shifts of electronic states can
be measured and from that the nuclear spin, electromagnetic moments,
and charge radius can be determined.
%the above could go to Experiments
Using this technique \cite{Geithner:2008}
made high-precision measurements of charge radii for neon
isotopes $^{17-22}$Ne. The charge radius for $^{17}$Ne,
$3.042(21)$~fm, was found to be the largest of all measured.
The results were compared with microscopic calculations
made in the fermionic molecular dynamics (FMD) framework
(\cite{Neff:2008}) which takes into account long-range
correlations, like halos or clusters. The FMD approach
was found to describe all the data very well, see Figure~\ref{fig:Halo_Ne-radii}.
The conclusion for $^{17}$Ne was that its large
radius is due to extended proton configuration with an $s^2$
component of about 40\%. In the next isotope, $^{18}$Ne, the
$s^2$ admixture is only about 15\%, and the radius is smaller.

\paragraph{\textit{$^{22}$Al and $^{27}$S}}

Recently a detailed $\beta$-delayed proton spectroscopy
for $^{22}$Si was undertaken by \cite{Lee:2020}. The properties of $\beta$ decay
branches to low-lying states in $^{22}$Al were measured
and compared with the corresponding mirror decay channels
of $^{22}$O. For the transition to the first excited $1^+$
state a very large asymmetry value, $\delta_{\beta} = 2.09(96)$
was found. In fact, it is the largest asymmetry ever reported
for low-lying states. Since $^{22}$Al has an $S_p$ value close to zero,
it could be a signature of the proton halo in this nucleus.

A similar observation, first indicated by \cite{Janiak:2017}
and later confirmed with higher accuracy by \cite{Sun:2019},
was made for the $\beta$ decay of $^{27}$S. The transitions
to the first excited states $3/2^+$ and $5/2^+$ in $^{27}$P,
compared to data for the mirror decay channels in the decay of $^{27}$Na,
revealed strong asymmetry with $\delta_{\beta}$ equal to $0.38(26)$ and
$0.48(18)$, respectively. This could point to the two-proton
halo in $^{27}$S: the two last protons in this nucleus
should occupy the $2s_{1/2}$ orbital, the separation energy is
$S_{2p}=728(78)$~keV, and it was among the first
proton-halo candidates predicted by \cite{Brown:1996}.

The mirror asymmetry in $\beta$
decay may result from abnormal structure of the initial
and/or the final nucleus. The $\beta$ daughter of $^{27}$S
is $^{27}$P, which is already an established proton-halo nucleus.
It is not excluded that in $^{22}$Si, which is at the proton dripline,
the valence protons have a substantial $2s_{1/2}$ component.
On the other hand, the mirror asymmetries reported for
$\beta$ decays of $^{22}$Si and $^{27}$S were calculated
using the available data for the mirror nuclei, $^{22}$O and $^{27}$Na.
As noted by \cite{Guadilla:2021} decay schemes for the latter seem to be
incomplete and must be remeasured with higher accuracy
before firm conclusions concerning mirror asymmetry are drawn.

It may turn out that all four nuclei, $^{27}$S, $^{27}$P, $^{22}$Si,
and $^{22}$Al exhibit features of proton halo, but to assess that,
certainly more experimental and theoretical studies are needed.
Actually, more investigations are required for all cases discussed
in this section to gain the complete understanding of the proton-halo
phenomenon, and for sure this will remain an important research
topic in the near future.

%=================================================================================
% Main author: CM
%\input{rp-process}
%=================================================================================
\section{Proton dripline nuclei and nucleosynthesis}
\label{sec:rp-process}

Nuclei in the proximity of the proton dripline play an important role also in energy
generation in high-temperature hydrogen burning (X-ray bursts, novae) and in the
nucleosynthesis of the heavier nuclei via the so-called rapid proton-capture (rp-) process.
The latter consists in a sequence of proton captures and $\beta$ decays responsible for the
burning of hydrogen into heavier elements under extreme conditions of temperature and density.
This network of nuclear reactions and decays proceeds close to the proton dripline and $N
\approx Z$ all the way to the $^{100}$Sn region, see the review by \cite{Schatz:1998} for
details. A limiting parameter for the reaction path followed by the rp-process is given by the
proton dripline, since proton capture is hindered by the negative separation energy and
$\beta$ decay therefore prevails. Such limit could be overcome by two-proton capture on the
last proton-bound isotone, but such process is not going to be dominant. If the reaction flow
reaches a proton-unbound nucleus or photo-disintegration reaction rates for an isotope in the
flow are inhibiting further proton-capture reactions, the reaction flow reaches a
``bottleneck'' and has to wait for the relatively slow $\beta$ decay to proceed further. The
``bottleneck'' nucleus is called a waiting point.

\begin{figure}[htbp]
\begin{center}
\includegraphics[width = 1.0\columnwidth]{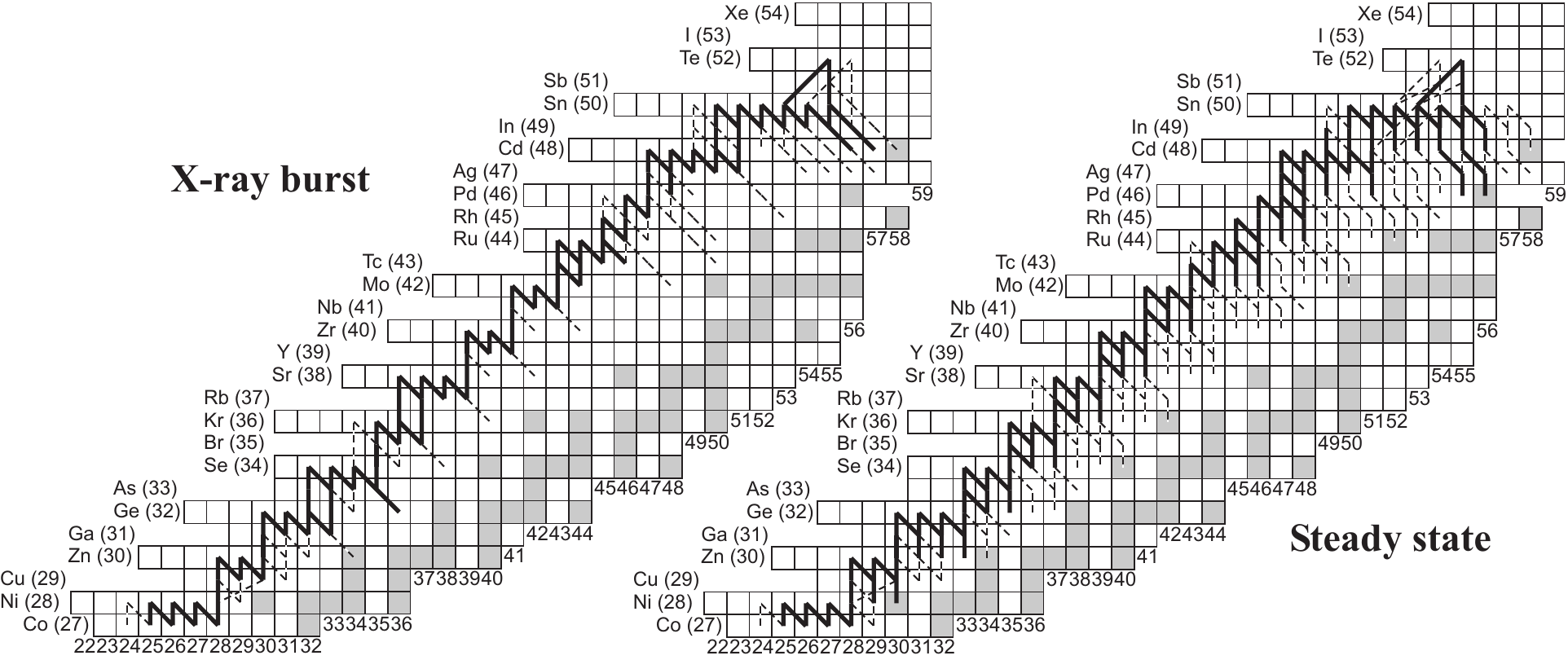}
\caption{Time-integrated reaction flow calculated for an X-ray burst and a steady-state
burning. Only the portion of the flow progressing beyond zinc and germanium is shown. Figure
from the work by \cite{Schatz:2001}. Copyright (2001) by the American Physical Society.}
\label{fig:rp_path}
\end{center}
\end{figure}

The flow, abundance patterns and energy output predicted by modelling of the rp-process are
extremely sensitive to the properties of the nuclei involved in the process. It was shown by
\cite{Schatz:1998} that the nuclear structure parameters that have the most impact on the
modelling are nuclear masses, since the main contribution to the shape of the rp-process path
is given by proton separation energies and $Q$-values. Also very important are $\beta$-decay
half-lives and branching ratios for delayed proton emission, which control the time structure
and final abundances, in particular those of the waiting points. Two-proton capture becomes
important in a high-density environment, since it can speed up the flow towards heavier
nuclei. It is therefore vital to determine the properties of the nuclei involved in the
rp-process, and to validate nuclear models that provide such properties for those isotopes
still not accessible in the laboratory.
Excited states in waiting point nuclei can also play a role in the rp-process flow, in
particular low-lying ones, which are populated at higher temperatures.
At temperatures of the order of 1-2~GK the contribution to the flow from the decay of excited
states cannot be ignored and may become of the same order of the contribution of the ground
state (\cite{Sarriguren:2011}).

A notable example of the impact of nuclear masses on the rp-process flow is illustrated by
calculations that predicted that, once the process reaches the tin isotopes with $A=$99-101,
it follows the isotopic chain towards less exotic nuclei and then gets ``trapped'' into a
cycle around tin-antimony-tellurium isotopes (\cite{Schatz:2001}), see
Figure~\ref{fig:rp_path}. The cycling is due to the fact that the tellurium isotopes involved
are $\alpha$ unbound and decay back to tin.
Whether the cycling around SnSbTe isotopes occurs or not, is entirely dependent on the nuclear
physics input to the calculations (binding energies, half-lives, branching ratios). The
occurrence or not of the cycle has important consequences for energy/light production and
composition of the ashes of the process. In fact, such a cycle, which happens late in the
process, produces helium, providing additional $\alpha$ particles that fuel reactions at later
times.
Intriguing is the impact of the proton separation energy in the most neutron-deficient
antimony isotopes on the termination of the rp-process. In fact, the precise determination of
the proton-separation energy in $^{105}$Sb and $^{106}$Sb led to a new scenario for the
rp-process path beyond $^{100}$Sn (\cite{Elomaa:2009,Mazzocchi:2007}), excluding almost
entirely the cycling around SnSbTe isotopes by confining its contribution to a level of 3\%.
The absence of strong cycling has the consequence of reducing production of helium at late
times.

Nuclear properties of exotic nuclei taking part into the rp-process, like proton and $\gamma$
widths ($\Gamma_p$ and $\Gamma_\gamma$) for excited states involved in the ($p,\gamma$) and
($\gamma,p$) reaction along the path, are often determined by means of indirect methods, given
that such nuclei are not easily accessible for direct reaction studies. A few examples of such
approaches can be found in the works by \cite{Langer:2014,Kennington:2021,Koldste:2013}

%=================================================================================
% Main author: CM & MP
%\input{Conclusions}
%=================================================================================
\section{Conclusions}
\label{sec:conclusions}

In this chapter an overview on the proton rich edge
of the chart of nuclei was presented and the main features and
phenomena which characterise these very proton-rich nuclei were introduced.
Of course, not all interesting results and research
ideas pertaining to the proton dripline region could be mentioned, and
illustrative examples which are most enticing in the authors opinion were selected.
The goal was to attract the reader to this domain of nuclear
physics, and to suggest links for further learning, rather than
to provide an exhaustive report.

One important aspect of proton dripline physics, is how much different research approaches are intertwined.
The same nucleus attracts attention for various reasons,
and different research methods, focused on different
facets of its structure, only in combination, yield the complete
picture. For example the $1/2^+$ state in $^{17}$F is the
subject to the Thomas-Ehrman shift, which leads to the proton
halo feature, which in turn affects the reaction rates of
astrophysical importance. The $^{100}$Sn and its region
is a ``holy grail" for the shell-model, a potential
reference point for $\alpha$ spectroscopy, and a key point
in the rp-process. Studies of isospin symmetry bring information
on nuclear forces but also deliver hints of anomalous
nuclear behaviour. Complementarity of methods is the way
to follow in this field.

Finally, a few remarks concerning
the future prospects of this field are worth noting.
The exploration of proton-rich nuclei continues and
the territory reached by experiment, marked in
Figures \ref{fig:Land_Chart_A} and \ref{fig:Land_Chart_B},
will soon expand. The new Facility for Rare Isotope Beams
(FRIB, see \cite{Sherrill:2018}) is just about to start operations.
According to estimates made by \cite{Neufcourt:2020}
the FRIB may shift the line of accessible isotopes by up
to 5 neutrons away from the current limit. In total,
about 120 new isotopes, between zirconium and uranium,
are expected to be delivered for studies.

When looking at the Figures~\ref{fig:Land_Chart_A} and \ref{fig:Land_Chart_B}
one question which suggests itself is how high in $Z$ number the
2\emph{p} radioactivity can occur and to how low $Z$ values the
observable \emph{p} emission will range. The progress in both
these lines of study requires increased production yields of
nuclei beyond the dripline. The search for light \emph{p}-emitters
will depend, in addition, on advances in fast techniques of
separation and detection, and on fast data acquisition methods which
rely to increasing extend on digital signal processing.

A more quantitative answer to the range of ground-state 2\emph{p}
emission was attempted by \cite{Olsen:2013,Olsen:2013b}.
Using advanced predictions of nuclear separation energies \cite{Erler:2012},
and simplified models of 2\emph{p} and $\alpha$ decays, they inquired
for which nuclei the 2\emph{p} half-life may be longer
then 100~ns, to be detectable using the in-flight technique,
and short enough to compete with other decay modes ($\beta^+$, $\alpha$).
The main conclusion was that conditions for 2\emph{p} radioactivity can
appear for all even-$Z$ elements between argon and tellurium.
Between xenon and lead, however, a different picture emerged.
Due to high Coulomb barrier, for the simultaneous emission of
two protons to be the dominant decay, one has to go so far
beyond the dripline, that a single proton becomes unbound.
In consequence, the two protons are emitted, but in sequence,
one after the other ($pp$). Above lead, $\alpha$ decay was found
to dominate totally. This general picture was confirmed by
a recent calculation based on the Bayesian Model Averaging method
by \cite{Neufcourt:2020}.

This kind of predictions are very
approximate, nevertheless it seems certain that when
going along the proton dripline towards lead, the region
between the dripline and the line of dominating proton emission,
becomes wider and wider. A spacious proton-rich \emph{terra incognita}
appears which could be addressed with classical nuclear spectroscopy.
To what extent it will be possible to explore this territory,
shall be seen when the new generation of radioactive beam facilities,
like FRIB at MSU,  FAIR at GSI, SPIRAL2 at GANIL, RAON in Korea,
will come into operation.

%=========================================================================
%\bibliographystyle{aps-nameyear}
%\bibliographystyle{aps}
%\bibliography{MPF_Handbook,ChM_Handbook}

\providecommand{\noopsort}[1]{}\providecommand{\singleletter}[1]{#1}%

\end{document}